\documentclass[journal]{IEEEtran}
\usepackage{array}
\usepackage{url}
\usepackage{amsmath}
\usepackage{amsthm}
\usepackage{amssymb}
\usepackage{graphicx}
\usepackage{cite}
\usepackage{subfig}

\newcommand{\noun}[1]{\textsc{#1}}
\providecommand{\tabularnewline}{\\}
\theoremstyle{plain}
\newtheorem{thm}{\protect\theoremname}
\theoremstyle{plain}
\newtheorem{prop}[thm]{\protect\propositionname}
\providecommand{\propositionname}{Proposition}
\providecommand{\theoremname}{Theorem}

\begin{document}

\title{Optimal Pricing of Internet of\\ Things: A Machine Learning Approach}

\author{Mohammad~Abu~Alsheikh,~\IEEEmembership{Member,~IEEE}, Dinh Thai Hoang,~\IEEEmembership{Member,~IEEE}, Dusit~Niyato,~\IEEEmembership{Fellow,~IEEE}, \\Derek Leong, Ping Wang,~\IEEEmembership{Senior~Member,~IEEE}, and Zhu Han,~\IEEEmembership{Fellow,~IEEE}

\thanks{M.~Abu~Alsheikh is with the Faculty of Science \& Technology, University of Canberra, ACT, Australia 2617 (email: mabualsh@ieee.org). D.~T.~Hoang is with the School of Electrical and Data Engineering, University of Technology Sydney, NSW, Australia 2007 (email: hoang.dinh@uts.edu.au). D.~Niyato is with the School of Computer Science and Engineering, Nanyang Technological University, Singapore 639798 (email: dniyato@ntu.edu.sg). P.~Wang is with the Department of Electrical Engineering \& Computer Science, York University, Canada M3J 1P3 (email: pingw@yorku.ca). Z.~Han is with the University of Houston, TX, USA 77004 (email: zhan2@uh.edu), and also with the Department of Computer Science and Engineering, Kyung Hee University, South Korea 446-701.}
}

\maketitle

\begin{abstract}
	Internet of things (IoT) produces massive data from devices embedded with sensors. The IoT data allows creating profitable services using machine learning. However, previous research does not address the problem of optimal pricing and bundling of machine learning-based IoT services. In this paper, we define the data value and service quality from a machine learning perspective. We present an IoT market model which consists of data vendors selling data to service providers, and service providers offering IoT services to customers. Then, we introduce optimal pricing schemes for the standalone and bundled selling of IoT services. In standalone service sales, the service provider optimizes the size of bought data and service subscription fee to maximize its profit. For service bundles, the subscription fee and data sizes of the grouped IoT services are optimized to maximize the total profit of cooperative service providers. We show that bundling IoT services maximizes the profit of service providers compared to the standalone selling. For profit sharing of bundled services, we apply the concepts of core and Shapley solutions from cooperative game theory as efficient and fair allocations of payoffs among the cooperative service providers in the bundling coalition.
\end{abstract} 
\section{Introduction}

Recent years have witnessed significant progress in using machine learning for solving challenging problems in the Internet of things~(IoT) including data caching~\cite{muller2017context,somuyiwa2018reinforcement}, activity recognition~\cite{xu2016personalized,alsheikh2016mobile}, channel estimation~\cite{prasad2014joint}, and security~\cite{xiao2016mobile,xiao2016phy}. Such machine learning-based solutions require massive datasets for system training. Trading IoT data among firms allows data vendors to make profit by offering their data to service providers over the Internet. Service providers use the bought data in creating and training advanced machine learning-based IoT services\footnote{For the rest of this paper, we use ``machine learning-based IoT services'' and ``services'' interchangeably.}, e.g., fraud detection, activity recognition, acoustic modeling, and medical diagnosis. However, economics of IoT services among service providers, data vendors, and customers is rarely studied in the literature. IoT market models and optimal pricing schemes are therefore required to ensure maximum profits of firms by achieving optimal utilization of the IoT resources and optimal subscription fees for services.

IoT services can be either sold separately or bundled and sold as one service package. In particular, multiple service providers, e.g., fraud detection and recommender systems, can cooperate to sell a bundled service at a discounted rate while sharing the resulting profit\footnote{Product bundling is a marketing strategy which is widely used by economists, e.g., a fast food meal consisting of a sandwich and a soft drink.}. Several major questions related to this service bundling process arise. \emph{Firstly, how is service quality defined and what are the optimal sizes of data that should be bought from data vendors? Secondly, should service providers cooperate to offer a bundled service instead of the standalone sales of services? Thirdly, once a cooperation is formed, how do cooperative service providers divide the resulting profit of a bundled service among themselves?}

This paper provides answers for the aforementioned questions by proposing IoT market models and optimal pricing schemes of selling machine learning-based IoT services separately or as bundled packages. The main objective is maximizing the profits of IoT service providers while providing bundled services to customers at discounted price. The key contributions of this paper can be summarized as follows:
\begin{itemize}
	\item From the service provider's perspectives, we define the service quality as a mapping between the bought data size and the resulting accuracy of machine learning algorithms.
	\item We develop an IoT market model for the standalone selling of services. We then propose a non-linear optimization problem with the objective of maximizing the profit of service providers. Specifically, a service provider will make decisions on the optimal data size to buy from data vendors and the optimal subscription fee that should be charged to service customers.
	\item We formulate and solve an IoT service bundling optimization to obtain the optimal bundle subscription fee and data sizes for each service in the bundle. Here, the optimization objective is to maximize the total profit of the cooperative providers inside the bundling coalition. We show that a bundled service increases the profit of service providers compared to the standalone sales of services. Moreover, bundled IoT services are favored by customers due to their discounted subscription fees compared to the standalone sales.
	\item We present a profit sharing scheme to divide the generated profit of a bundled IoT service among the cooperative providers based on their contributions to the bundle profit. We apply the concepts of core solution and Shapley value to find the payoff allocations to providers.
\end{itemize}
Additionally, we provide closed-form solutions for each optimization problem in this paper. Unlike iterative solutions of nonlinear problems with inequality constraints, e.g., interior point methods~\cite{boyd2004convex}, closed-form solutions can be evaluated in a finite time and are computationally efficient.

The rest of this paper is organized as follows. Section~\ref{sec:related_work} presents the related work. Section~\ref{sec:system_model} discusses the IoT market model and assumptions, and machine learning-based model is presented as a measure of service qualities. Then, optimization problems for profit maximizing are derived in Section~\ref{sec:separate_selling} for the standalone services and in Section~\ref{sec:service_bundling} for the bundled IoT services. Section~\ref{sec:revenue_sharing } presents a model for payoff allocation and sharing among service providers in service bundling. Section~\ref{sec:numerical_results} discusses the experimental evaluation results. The paper is finally concluded in Section~\ref{sec:conclusion}.

\section{Related work\label{sec:related_work}}

A survey of $3,000$ employees from $100$ countries reports that successful firms apply machine learning five times more than low-performing ones~\cite{lavalle2011big}. This clearly shows the importance of adopting the recent advances in machine learning for generating revenues and meeting the market demands on intelligent IoT systems. Moreover, market models and pricing strategies are integral for maximizing the profit of selling products and services including wired and wireless network access~\cite{sen2013survey,seregina2016design, zhang2019hybrid}, cloud computing~\cite{khodak2018learning}, and mobile crowdsensing~\cite{jiang2017scalable,xu2019ilocus}, just to name a few.

\subsection{Economics of Information Goods}

Pricing information goods, e.g.,~software products and movies, is a well-studied problem in the literature. In~\cite{jain2002pricing}, three pricing schemes of information goods were presented, namely, connection time-based, search-based, and subscription-based pricing. It has been argued that connection time-based pricing is less profitable than the other schemes for highly skilled users. The authors of~\cite{sundararajan2004nonlinear} presented a nonlinear pricing scheme of information goods based on the customer usage, i.e., defining a fixed price to each usage level. In~\cite{balasubramanian2015pricing}, the authors discussed the pay-per-use and unlimited subscription of information goods. It is shown that a maximum profit is achieved through competitive pricing. The authors in~\cite{bakos1999bundling} discussed the benefits of bundling information goods. It is shown that an accurate estimation of the user behavior can be achieved for bundled goods compared to the individual sales of products.

Pricing and bundling of IoT services is more challenging than those of classical information goods, as the quality of information goods can be easily determined, e.g.,~the quality of a software is defined based on its supported features, while the casting and genre of a movie define its price. On the contrary, the quality of IoT services cannot be directly measured.

\subsection{Economics of Query-based Data Services}

Query-based data services extract data from structured databases and visualize them according to customer requests. Simple query pricing models can be preferable by buyers, but applying a complex pricing model increases the revenues of the content provider~\cite{balazinska2011data}. The authors in~\cite{koutris2015query} presented a query-based pricing scheme of data. The price of a query, i.e., a question, is defined based on the number of data views required to provide the answer. In~\cite{li2014theory}, the authors presented a pricing method in privacy-preserving query systems. This differential privacy pricing aims at maximizing the accuracy of queries while maintaining the privacy of the data owners. The privacy level is defined by the query buyers. The authors in~\cite{lin2014arbitrage} proposed a query pricing scheme with arbitrage-prevention. This prevents buyers from combining simple and cheap queries to achieve the same results of a complex and expensive query. Such existing query-based schemes are restricted by design to structured and relational data, and they do not address unstructured data, e.g.,~IoT data, which is the dominant form of data in real-world settings.

\subsection{Incentive Mechanisms for Crowdsensing}

Incentive mechanisms are required for maintaining high user participation and providing fair reward allocation. The authors in~\cite{yang2015incentive} presented a user-centric incentive mechanism formulated as a reverse auction, and a platform-centric incentive scheme modeled as a Stackelberg game. The user-centric mechanism enables the participants to compete for a higher reward. In the platform-centric mechanism, the service acts as the game leader and announces the reward of completing the data collection task. The participants are the game followers and set their sensing time in order to maximize the received reward. Likewise, the authors in~\cite{singla2013truthful} modeled the crowdsensing problem as an online auction, while assuming that the arrival of participants has a stochastic distribution. In~\cite{luo2016incentive}, the reward allocation problem was modeled as an all-pay auction such that the participant with the highest contribution receives the full reward.

This paper is fundamentally different from all past works. First, the problem of defining the quality of service in IoT has not been previously addressed. Second, existing works have not considered the data demand and pricing in IoT services when machine learning is heavily utilized. Third, none of the existing works has considered the problem of cooperation among service providers to form a bundled service which is an effective strategy for profit maximization. This paper addresses these limitations and presents a market model, bundling strategies, and optimal pricing schemes of IoT services. The novel data-driven optimizations of this paper allow achieving the maximum profits from offering IoT services to customers.

\section{System Model\label{sec:system_model}}

In this section, we first briefly introduce the concept of machine learning-based IoT services and give real world examples of IoT services where optimal pricing and bundling are required. Then, we present a method for defining the value of data from a machine learning perspective.

\subsection{Machine Learning-Based IoT Services}

\begin{figure}
	\begin{centering}
		\includegraphics[width=1.0\columnwidth,trim=1cm 1cm 1cm 2cm]{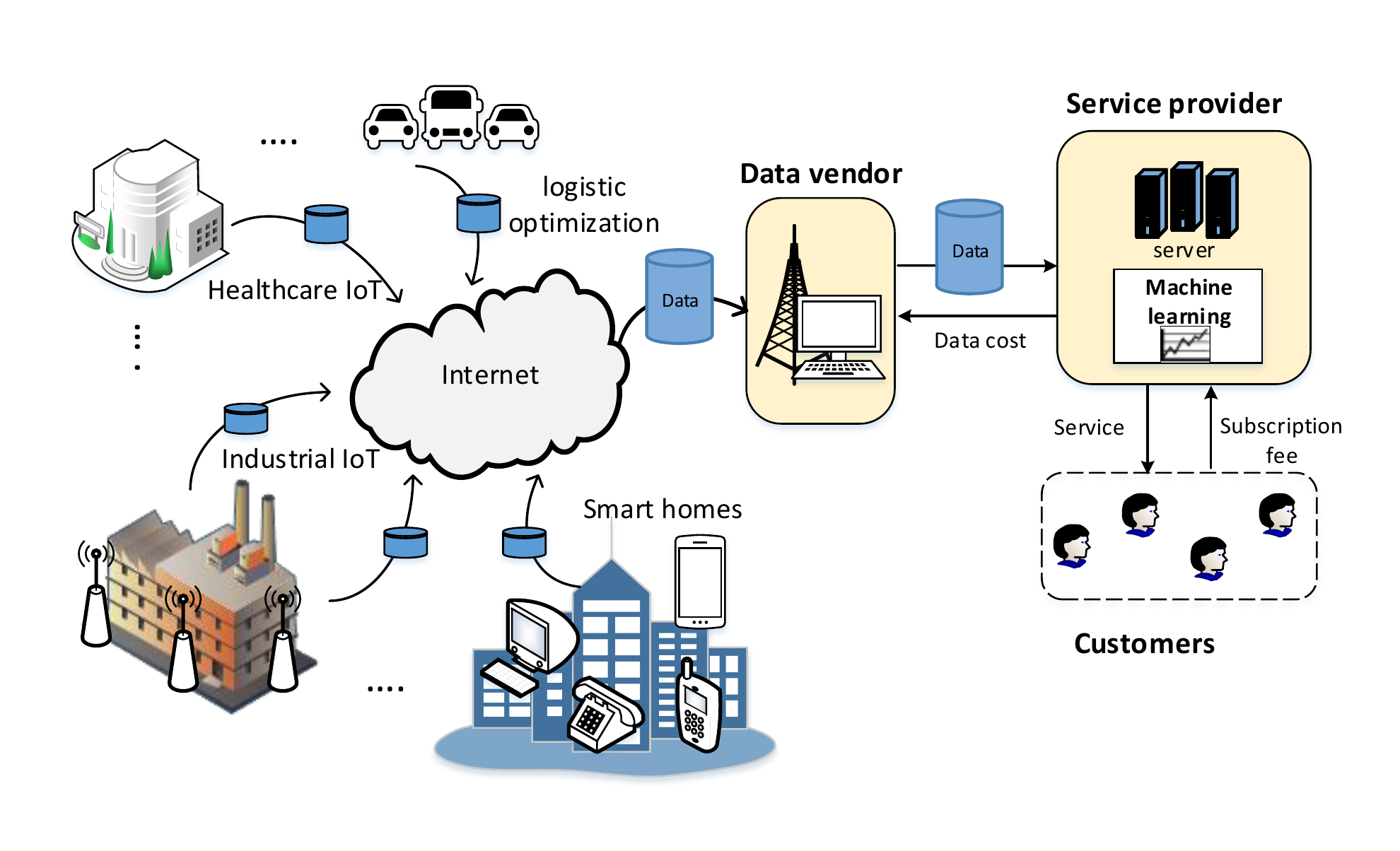}
		\par\end{centering}
	
	\caption{System model of machine learning-based IoT services.\label{fig:market_models}}
\end{figure}

We consider the IoT service architecture shown in Figure~\ref{fig:market_models}. An IoT service is typically composed of a data vendor, service provider, and service customers.
\begin{itemize}
	\item \emph{Data vendors}: The IoT data is firstly collected, stored and filtered by a data vendor. The data can be generated by different devices and technologies, e.g., sensor nodes, IoT gadgets, and smart phones. In addition to the deployment cost, the process of data gathering also requires costly human intervention for data annotation, validation, and preprocessing, e.g., anomaly detection and missing data imputation. Accordingly, the data vendor charges data buyers with a price. The cost of one data unit\footnote{We assume that a data unit includes one percent of the full dataset samples.} is denoted by $c$.
	\item \emph{Service provider}: The raw data owned by data vendors is unprofitable unless suitable analytics tools are applied. A service provider is a business entity which buys data from one or more data vendors, uses machine learning tools, and offers a service to customers willing to pay a subscription fee. For profit maximization, a rational service provider decides the data size $n$ (in data units) to buy from data sources and the subscription fee $p_{s}$ to charge for his service.
	\item \emph{Service customers}: We assume that there are a total of $M$ customers. Each customer is an independent entity who decides whether to subscribe to a service provider based on his willingness-to-pay $\theta$ for the service and the subscription fee $p_{s}$ set by the provider. $\theta$ defines the maximum subscription fee that a customer can pay for a service based on his evaluation and need for the service.
\end{itemize}
This IoT market model is useful in many service-oriented architectures.
\begin{itemize}
	\item A \emph{data marketplace} is an online store where entities can trade data. machine learning, e.g., machine learning algorithms, can be applied on the data to generate prediction models. Datasets of various types are offered and exchanged as assets. Examples of data markets include Azure Marketplace\footnote{\url{https://www.datamarket.azure.com}}, Qlik DataMarket\footnote{\url{https://www.qlik.com}}, and Infochimps\footnote{\url{http://www.infochimps.com}}.
	\item \emph{Crowdsourcing services} also require optimized market models for data exchange\emph{. Placemeter}\footnote{\url{https://www.placemeter.com}}, for example, provides real-time information of pedestrian and vehicular movement in cities and urban areas. Placemeter allows users to upload videos of streets and public areas and they are paid back based on the video quality. Computer vision algorithms are used by Placemeter to extract information from real-time data.
	\item \emph{IoT services} is a new trend of IoT platforms which are designated to visualize and trade IoT data. Thingful\footnote{\url{https://www.thingful.net}} is an example of these platforms which allows IoT vendors and owners to visualize geolocations of connected devices around the world. Likewise, health care systems, such as PatientsLikeMe\footnote{\url{https://www.patientslikeme.com}}, sell rich medical data which can be collected with IoT gadgets and other smart technologies.
\end{itemize}
The list of symbols used in this paper are summarized in Table~\ref{tab:list_symbols}.

\begin{table}
	\caption{List of frequently used notations and symbols throughout the paper.\label{tab:list_symbols}}

	\centering{}%
	\begin{tabular}{|c|>{}p{0.6\columnwidth}|}
		\hline 
		\textbf{\noun{Symbol}} & \textbf{\noun{Definition}}\tabularnewline
		\hline 
		\hline 
		$\mathcal{D}$ & Dataset\tabularnewline
		\hline 
		$n$ & Requested data size\tabularnewline
		\hline 
		$c$ & Cost of one data unit \tabularnewline
		\hline 
		$M$ & Number of customers\tabularnewline
		\hline 
		$p_{s}$ & Subscription fee\tabularnewline
		\hline 
		$\theta$ & Willingness-to-pay (reservation price) for a service by customers\tabularnewline
		\hline 
		$F\left(\cdot\right)$ & Profit function of a service provider\tabularnewline
		\hline 
		$\mathbf{H}$ & Hessian matrix of $F\left(\cdot\right)$ \tabularnewline
		\hline 
		$\Delta_{i}$ & $i$th-order principal minor of $\boldsymbol{H}$ \tabularnewline
		\hline 
		$\mathcal{L}\left(\cdot\right)$ & Lagrangian dual problem \tabularnewline
		\hline 
		$\lambda_{i}$ & Lagrangian multiplier\tabularnewline
		\hline 
		$p_{b}$ & Bundle subscription fee\tabularnewline
		\hline 
		$q(\cdot)$ & Service quality function\tabularnewline
		\hline 
		$\alpha$ & Fitting parameters of $q(\cdot)$ \tabularnewline
		\hline 
		$\mathcal{C}$ & The core solution set\tabularnewline
		\hline 
		$\mathcal{K}$ & Bundling coalition\tabularnewline
		\hline 
		$F_{\mathcal{K}}\left(\cdot\right)$ & Profit function of a bundling coalition $\mathcal{K}$\tabularnewline
		\hline 
		$\mathbf{H}_{\mathcal{K}}$ & Hessian matrix of $F_{\mathcal{K}}\left(\cdot\right)$ \tabularnewline
		\hline
		$\mathcal{L}_{\mathcal{K}}\left(\cdot\right)$ & Lagrangian dual function of a bundling coalition $\mathcal{K}$\tabularnewline
		\hline 
		$\varphi_{k}$ & Payoff allocation for provider $k\in\mathcal{K}$ in the core solution\tabularnewline
		\hline 
		$\eta_{k}$ & Payoff allocation for provider $k\in\mathcal{K}$ by the Shapley solution\tabularnewline
		\hline 
	\end{tabular}
\end{table}

\subsection{Machine Learning in IoT}

\begin{figure}
	\begin{centering}
		\includegraphics[width=0.9\columnwidth]{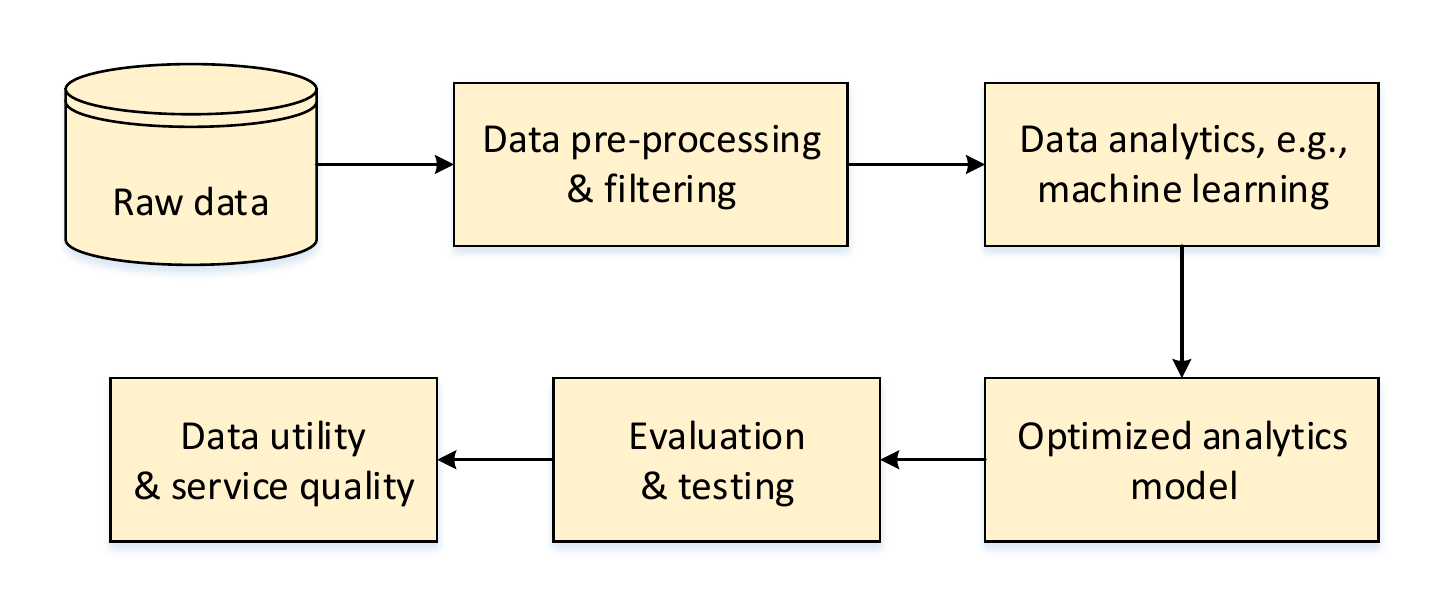}
		\par\end{centering}
	
	\caption{Applying machine learning in IoT services.\label{fig:data_mining_analytics}}
\end{figure}

As shown in Figure~\ref{fig:data_mining_analytics}, consider a training data $\mathcal{D}=\left\{ \mathbf{x}_{i},y_{i}\right\} _{i=1}^{B}$ which includes $B$ tuples of connected feature set $\mathbf{x}_{i}\in\mathbb{R}^{R}$ and a class label $y_{i}\in\mathbb{R}$. $y_{i}$ is only present in supervised learning, e.g., classification and regression problems. Based on the problem at hand, a feature set $\mathbf{x}_{i}$ contains sensing data of $R$ features such as images in vision problems, audio in acoustic modeling, text in document classification, and temperature values in weather forecasting. Generally, a machine learning optimization problem adheres to the following general formula:
\begin{equation}
\begin{aligned}\underset{\Psi}{\text{minimize }} & e\left(\mathcal{D};\Psi\right)+\gamma g\left(\Psi\right),\end{aligned}
\label{eq:ml_objective}
\end{equation}
where $e\left(\cdotp\right)$ is a learning objective function designed to fit a model $\Psi$ to historical data $D$, $g\left(\cdotp\right)$ is a regularization term, $\gamma$ is a weighted hyper-parameter. For example, a deep network~\cite{Goodfellow-et-al-2016-Book} has a learning function which is defined as follows:

\begin{equation}
\underbrace{\frac{1}{B}\sum_{i=1}^{B}\left(\frac{1}{2}\left\Vert h_{\Psi}\left(\mathbf{x}_{i}\right)-y_{i}\right\Vert ^{2}\right)}_{e\left(\mathcal{D};\Psi\right)}+\frac{\gamma}{2}\underbrace{\sum_{l=1}^{L-1}\sum_{i=1}^{s_{l}}\sum_{j=1}^{s_{l+1}}\left(\Psi_{ji}^{(l)}\right)^{2}}_{g\left(\Psi\right)},
\end{equation}
where $L$ is the number of layers in the deep model, $s_{l}$ is the number of neurons at layer $l$, $h_{\Psi}\left(\mathbf{x}_{i}\right)$ is the deep model prediction for input $\mathbf{x}_{i}$. The first term is an average sum-of-squared errors, the second term is a weight decay regularization which includes the summation modeling parameters. After the model is trained using the training data, it is tested to define the accuracy using the unseen testing data.

\subsection{Service Quality}

One of the key barriers in the development of data marketplaces is defining data quality and value to the potential service providers. In this section, we define the data quality based on the performance of machine learning models trained using the data. 

Customers do not always subscribe to the cheapest IoT services. Instead they infer the subscription fee and service quality. More data is important to increase the accuracy of IoT services~\cite{halevy2009unreasonable,domingos2012few}. However, this accuracy gain increases the cost of bought data. From the service provider's perspectives, we define the quality function $q(\cdot)$ of a service as a mapping between the bought data size to the resulting accuracy of the machine learning. The quality of the service is equal to the utility of bought data from the service provider's perspectives. For example, machine learning in intrusion detection systems by energy-constrained sensor networks is aimed for locating and identifying intruders~\cite{abu2014machine}. The cost of detecting an intruder is a direct reflection of the value of data.

We introduce the service quality function $q(\cdot)$ to meet the following empirical assumptions:
\begin{itemize}
	\item $q(\cdot)$ is an increasing function such that $q'(\cdot)>0$. This assumption is intuitive as more data improves the accuracy of the analytics and quality of the service.
	\item $q(\cdot)$ has a decreasing marginal utility such that $q''(\cdot)\leq0$. This assumption reflects the empirical accuracy of machine learning models.
\end{itemize}
In our data market and pricing framework, $q(\cdot)$ is defined as follows:
\begin{equation}
q(n;\mathbf{\mathbf{\alpha}})=\alpha_{1}-\alpha_{2}\exp(-\alpha_{3}n),\label{eq:data_quality}
\end{equation}
where $n$ is the data size and $\mathbf{\mathbf{\alpha}}=\left[\begin{array}{ccc} \alpha_{1} & \alpha_{2} & \alpha_{3}\end{array}\right]$ is a tuple of three fitting parameters. To find the fitting parameters $\mathbf{\alpha}$, we vary the size of raw data $\mathcal{D}$ used to fit the model $\Psi$. Specifically, a series of $O$ experimentation points $\left(n^{(1)},\varepsilon^{(1)}\right),\ldots,\left(n^{(j)},\varepsilon^{(j)}\right),\ldots,\left(n^{(O)},\varepsilon^{(O)}\right)$ is performed, where $\varepsilon^{(j)}$ is the service accuracy resulting from a data size of $n^{(j)}$ with $n^{(j)}>n^{(j+1)}$. $q(n;\mathbf{\mathbf{\alpha}})$ is then found by applying nonlinear least squares for minimizing the sum of squared errors as follows:
\begin{equation}
\text{\ensuremath{\underset{\alpha}{\text{minimize }}}}\frac{1}{O}\sum_{j=1}^{O}\left\Vert \varepsilon^{(j)}-q\left(n^{(j)};\mathbf{\mathbf{\alpha}}\right)\right\Vert ^{2}\label{eq:param_fitting}.
\end{equation}
The parameter fitting problem in (\ref{eq:param_fitting}) can be solved iteratively to find the fitting parameter $\alpha$~\cite{strutz2010data}. $O$ can be estimated before transmitting the data from the data vendor to the service provider through a third party broker which charges a broker fee. In real-world IoT services, the broker can also manage the financial transactions, service performance, and service delivery, e.g.,~measuring the compliance with the service-level agreement~(SLA).

In Section~\ref{sub:exp_data_utility}, we present extensive experiments that show the validity of (\ref{eq:data_quality}) in fitting the real accuracy of machine learning-based IoT  services. For simplified notations, we use $q$ instead of $q(n;\mathbf{\mathbf{\alpha}})$ in the rest of this paper.

\section{IoT Market Model and Optimal Pricing of Standalone Services\label{sec:separate_selling}}

In this section, we first develop an IoT market model of selling standalone IoT services, and we formulate an optimization problem to maximize the profit of a service provider. Then, the closed-form solutions for the optimal subscription fee and requested data size are provided.

We consider the separate sales of a monopolist service provider providing a service to a set of $M$ potential customers as shown in Figure~\ref{fig:market_models}. Each customer has a different willingness-to-pay $\theta$ for the data service. If the willingness-to-pay of a particular customer is higher than or equal to the subscription fee $p_{s}$ weighted by the service quality $q$ such that $\theta q\geqslant p_{s}$, that customer will subscribe to the service. This indicates that the willingness-to-pay value depends on the customer evaluation of the service as well as its quality, i.e., high quality services will attract more customers. Based on our real-world customer surveys (Figure~\ref{fig:wtp_surveys}), we show that $\theta$ follows a uniform distribution. Then, the profit of the service provider is found as follows:
\begin{eqnarray}
F\left(p_{s},n\right) & = & \underbrace{Mp_{s}\mathbb{P}\left(\theta q\geqslant p_{s}\right)}_{\text{revenue}}-\underbrace{nc}_{\text{data cost}}\label{eq:seperate_profit}\\
& = & Mp_{s}\left(1-\frac{p_{s}}{q}\right)-nc
\end{eqnarray}
where $q=\alpha_{1}-\alpha_{2}\exp\left(-\alpha_{3}n\right)$ is the service quality function. The first term of (\ref{eq:seperate_profit}) is the revenue of offering the service to the customers with the subscription fee $p_{s}$. As $\theta$ is assumed to follow a uniform distribution, the expression $\mathbb{P}\left(\theta q\geqslant p_{s}\right)=1-\frac{p_{s}}{q}$ defines the probability of subscribing to the service by the customers. The second term of (\ref{eq:seperate_profit}) is the total data price paid to the data vendor which is equal to the requested data size multiplied by the data unit cost. The marginal cost of running the service is negligible.

We next formulate a profit maximization optimization based on the profit function in (\ref{eq:seperate_profit}) for the service provider to decide their optimal requested data size and subscription fee.

\subsection{Profit Maximization of Service Providers}

The service provider is assumed to be rational and is interested in maximizing his individual profit. A non-linear optimization problem can be formulated to choose the optimal data size $n^{*}$ that should be bought from the data vendor and optimal subscription fee $p_{s}^{*}$ to be charged to the customers as follows:

\begin{equation}
\begin{aligned}\underset{n,p_{s}}{\text{maximize }} & F\left(p_{s},n\right)=Mp_{s}\left(1-\frac{p_{s}}{q}\right)-nc\\
\text{subject to } & C1:p_{s}\geq0,\\
& C2:n\geq0.
\end{aligned}
\label{eq:seperate_selling_problem}
\end{equation}
The objective function of (\ref{eq:seperate_selling_problem}) maximizes the profit of the service provider. The two constraints $C1$ and $C2$ are required to ensure non-negative solutions for the subscription fee and requested data size.
\begin{prop}
	$F\left(p_{s},n\right)$ is concave, and hence the closed-form solution
	of the optimization problem $\underset{p_{s},n}{\text{maximize }}F\left(p_{s},n\right)$
	is globally optimal.
	\label{prop:seperate_selling_problem}
\end{prop}

\begin{IEEEproof}
	We use Sylvester's criterion of twice differentiable functions to prove that the Hessian matrix $\boldsymbol{H}$ of $F\left(p_{s},n\right)$ is negative semidefinite, and hence the concavity of $F\left(p_{s},n\right)$~\cite{chong2013introduction}. Suppose that the principal minors of $\boldsymbol{H}$ are denoted as $\left\{ \Delta_{i}\right\} _{i=1,2}$, where $i$ is the order of the principal minors. Based on Sylvester’s criterion, $\boldsymbol{H}$  is  negative semidefinite if and only if the condition $\left(-1\right)^{i}\Delta_{i}\geq0$ holds, i.e.,~every odd-order principal minor is non-positive and every even-order principal minor is non-negative. The Hessian matrix $\boldsymbol{H}$ of $F\left(p_{s},n\right)$ is obtained as shown in (\ref{eq:hessian_seperate}) at the top of the next page. The first-order principal minors of $\boldsymbol{H}$ are derived as follows:
	\begin{equation}
	\Delta_{1,1} = \frac{-2M}{q}\leq0,
	\end{equation}
	\begin{multline}
	\Delta_{1,2} = \frac{-2M\alpha_{2}^{2}\alpha_{3}^{2}p_{s}^{2}\exp\left(-2\alpha_{3}n\right)}{q^{3}}
	\\-\frac{M\alpha_{2}\alpha_{3}^{2}p_{s}^{2}\exp\left(-\alpha_{3}n\right)}{q^{2}}\leq0.
	\end{multline}
	The principal minor of order two is derived as follows:
	\begin{equation}
	\Delta_{2,1}=\frac{2\alpha_{2}\left(M\alpha_{3}p_{s}\right)^{2}}{q^{3}}\geq0.
	\end{equation}
	Accordingly, Sylvester's criterion is satisfied. $\boldsymbol{H}$ is negative semidefinite and $F\left(p_{s},n\right)$ is concave.
\end{IEEEproof}

\begin{figure*}
	\begin{equation}
	\mathbf{H}=\left[\begin{array}{cc}
	\dfrac{-2M}{q}&
	\dfrac{2M\alpha_{2}\alpha_{3}p_s\exp\left(-\alpha_{3}n\right)}{q^2}\\
	\dfrac{2M\alpha_{2}\alpha_{3}p_s\exp\left(-\alpha_{3}n\right)}{q^2} &
	\dfrac{-2M\alpha_{2}^2\alpha_{3}^2p_s^2\exp\left(-2\alpha_{3}n\right)}{q^3} - \dfrac{M\alpha_{2}\alpha_{3}^2p_s^2\exp\left(-\alpha_{3}n\right)}{q2}
	\end{array}\right].
	\label{eq:hessian_seperate}
	\end{equation}
	\noindent\rule{\textwidth}{1pt}
\end{figure*}

We next apply the Karush\textendash Kuhn\textendash Tucker~(KKT) conditions to find the optimal solutions of (\ref{eq:seperate_selling_problem}). The KKT conditions are necessary and sufficient conditions for optimality when applied to concave functions~\cite{boyd2004convex}.

\subsection{Optimal Subscription Fee and Requested Data Size}

Converting the constrained profit maximization problem (\ref{eq:seperate_selling_problem}) into a minimization problem and formulating the Lagrangian dual problem results in the following unconstrained Lagrange dual function: 
\begin{equation}
\mathcal{L}\left(\left\{ p_{s},n\right\} ,\lambda_{1},\lambda_{2}\right)=-Mp_{s}\left(1-\frac{p_{s}}{q}\right)+nc-\lambda_{1}p_{s}-\lambda_{2}n,\label{eq:seperate_selling_dual}
\end{equation}
where $\lambda_{1}$ and $\lambda_{2}$ are Lagrange multipliers for the constraints $C1$ and $C2$, respectively. The first derivatives of (\ref{eq:seperate_selling_dual}) with respect to $p_{s}$ and $n$ are as follows:

\begin{flalign}
\frac{\partial\mathcal{L}\left(\cdot\right)}{\partial p_{s}} & =M\left(\frac{p_{s}}{q}-1\right)+\frac{Mp_{s}}{q}-\lambda_{1},\\
\frac{\partial\mathcal{L}\left(\cdot\right)}{\partial n} & =c-\lambda_{2}-\frac{M\alpha_{2}\alpha_{3}p_{s}^{2}\exp\left(-\alpha_{3}n\right)}{q^{2}},
\end{flalign}
where $\lambda_{1}=0$ and $\lambda_{2}=0$ (no active constraints). Setting both derivatives to zero, the optimal closed-form solutions are found as follows:

\begin{flalign}
p_{s}^{*} & =\frac{M\alpha_{1}\alpha_{3}-4c}{2M\alpha_{3}},\label{eq:eq:seperate_sol_ps}\\
n^{*} & =\frac{1}{\alpha_{3}}\log\left(\frac{M\alpha_{2}\alpha_{3}}{4c}\right),\label{eq:seperate_sol_n}
\end{flalign}
where the condition $M\alpha_{1}\alpha_{3}\geq4c$ must hold to ensure positive values of $n^{*}$ and $p_{s}^{*}$. Accordingly, the maximum profit $F\left(p_{s}^{*},n^{*}\right)$ of the provider is found by substituting the optimal values of $n^{*}$ and $p_{s}^{*}$ in (\ref{eq:seperate_profit}).
\section{IoT Market Model and Optimal Pricing of Bundled Services\label{sec:service_bundling}}

In this section, we first present a market model where bundling is used by two cooperative service providers (denoted as Service~1 and Service~2). We denote the bundling coalition as $\mathcal{K}$. The two services are grouped together in one package and sold at a discounted subscription fee, i.e., subscribing to the two services separately costs more than the bundled service. It is important to note that customers can still subscribe to services separately. Then, we develop an optimization problem to select the optimal bundle subscription fee and requested data sizes by both providers.

\begin{figure}
	\begin{centering}
		\includegraphics[width=1.0\columnwidth,trim=0cm 0.5cm 0cm 0cm]{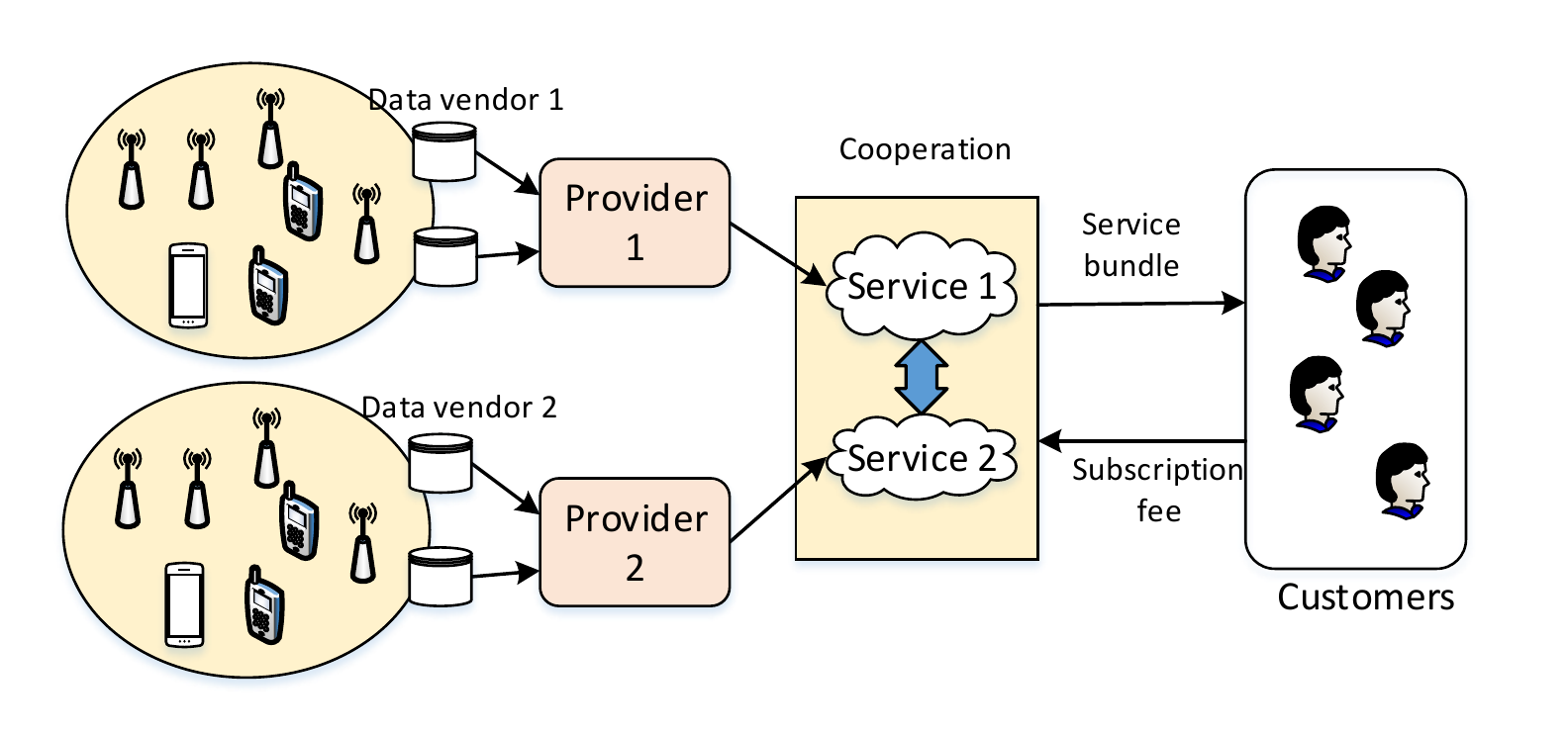}
		\par\end{centering}
	\caption{System model of offering IoT services as one bundle.\label{fig:service_bundling}}
\end{figure}

\subsection{Profit Maximization of Bundled Services}

\begin{figure}
	\begin{centering}
		\includegraphics[width=1.0\columnwidth,trim=0.5cm 0cm 1cm 0cm]{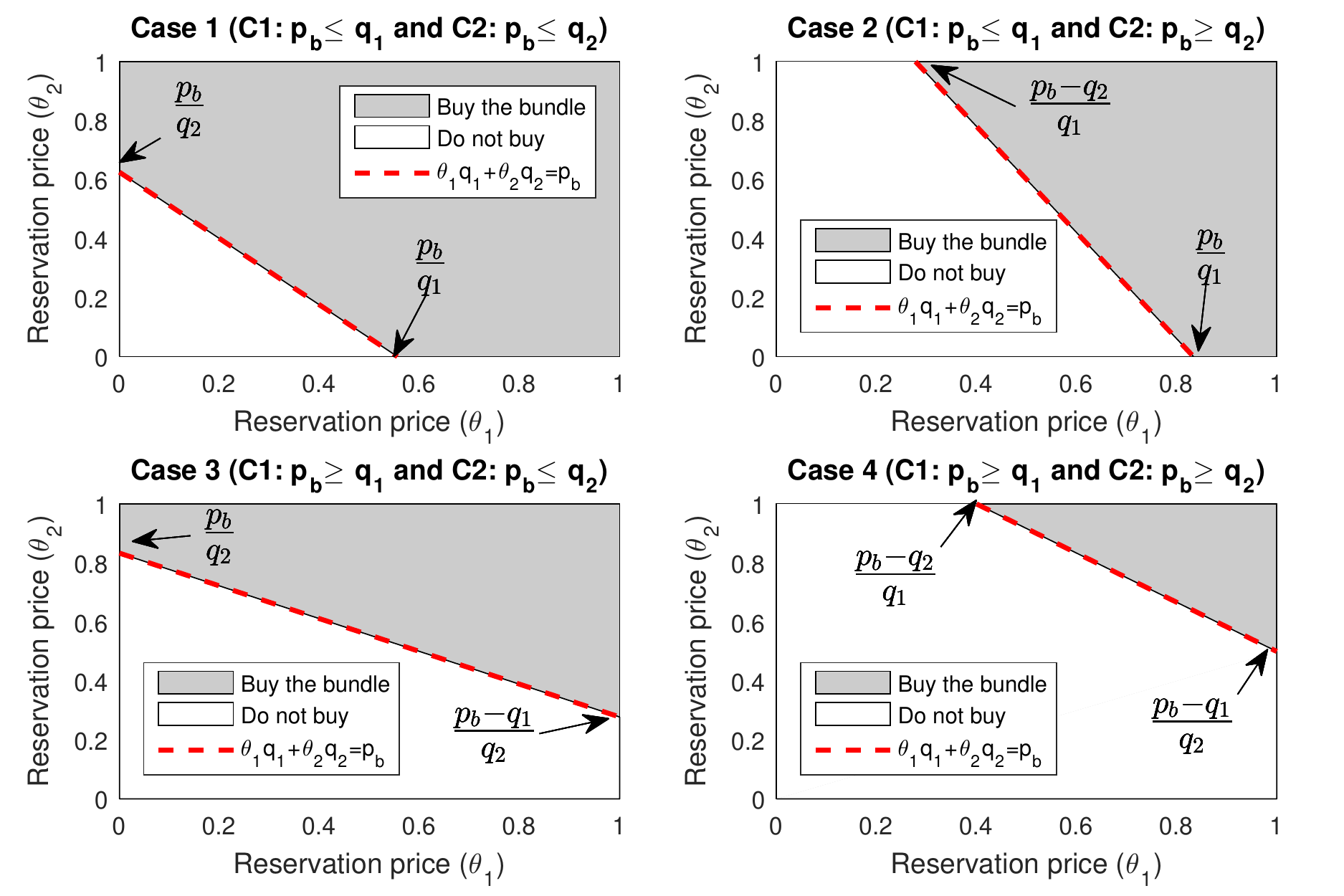}
		\par\end{centering}
	
	\caption{The four demand cases on a bundled IoT service.\label{fig:bundling_cases}}
\end{figure}

Consider two service providers that form a coalition $\mathcal{K}$ to provide a bundled service to a base of $M$ customers as shown in Figure~\ref{fig:service_bundling}. The subscription fee of the bundle is denoted as $p_{b}$. The qualities of Services~1 and 2 are denoted as $q_{1}$ and $q_{2}$, respectively. The data unit costs paid by Services~1 and 2 are $c_{1}$ and $c_{2}$, respectively. We define the reservation prices $\theta_{1}$ and $\theta_{2}$ as the willingness of the customers to pay for Services~1 and 2 in the bundle. Specifically, a customer will decide to subscribe to a bundled service if the following relation hold
\begin{equation}
\theta_{1}q_{1}+\theta_{2}q_{2}\geq p_{b}\label{eq:bundle_decision}
\end{equation}
which indicates that the customer evaluation of the bundled service is higher than its subscription fee. The profit optimization problem of a bundled service is then defined as follows:
\begin{equation}
\begin{aligned}\underset{p_{b},n_{1},n_{2}}{\text{maximize }} & F_{\mathcal{K}}\left(p_{b},n_{1},n_{2}\right)=\\
& \underbrace{Mp_{b}\mathbb{P}\left(\theta_{1}q_{1}+\theta_{2}q_{2}\geq p_{b}\right)}_{\text{revenue}}-\underbrace{n_{1}c_{1}-n_{2}c_{2}}_{\text{data cost}}\\
\text{subject to } & C1,C2,\\
& C3:p_{b}\geq0,\\
& C4:n_{1}\geq0,\\
& C5:n_{2}\geq0,
\end{aligned}
\label{eq:bundling_problem}
\end{equation}
where $\mathbb{P}\left(\theta_{1}q_{1}+\theta_{2}q_{2}\geq p_{b}\right)$ is the demand on the bundle service by customers, i.e., the probability that a customer will decide to subscribe to a bundled service. The objective function of (\ref{eq:bundling_problem}) maximizes the total bundling profit of cooperative service providers which is equal to the sale revenue minus the paid data cost. $C3$, $C4$, and $C5$ are required to ensure positive solution values for $p_{b}$, $n_{1}$, and $n_{2}$, respectively. The constraints $C1$ and $C2$ are the optimization constraints which depend on the bundle price and service quality. In particular, there are 4 possible demand patterns (defining the optimization constraints $C1$ and $C2$) where (\ref{eq:bundle_decision}) holds as demonstrated by the shaded areas in Figure~\ref{fig:bundling_cases}. Specifically, $\mathbb{P}\left(\theta_{1}q_{1}+\theta_{2}q_{2}\geq p_{b}\right)$ can be found for each case as follows:
\begin{flalign}
\text{Case 1: } & 1-0.5\frac{p_{b}}{q_{1}}\frac{p_{b}}{q_{2}},\label{eq:demand_case_1}\\
\text{Case 2: } & 0.5Mp_{b}\left(1-\frac{p_{b}}{q_{1}}+1-\frac{\left(p_{b}-q_{2}\right)}{q_{1}}\right),\label{eq:demand_case_2}\\
\text{Case 3: } & 0.5Mp_{b}\left(1-\frac{p_{b}}{q_{2}}+1-\frac{\left(p_{b}-q_{1}\right)}{q_{2}}\right),\label{eq:demand_case_3}\\
\text{Case 4: } & 0.5Mp_{b}\left(1-\frac{\left(p_{b}-q_{1}\right)}{q_{2}}\right)\left(1-\frac{\left(p_{b}-q_{2}\right)}{q_{1}}\right).\label{eq:demand_case_4}
\end{flalign}
(\ref{eq:demand_case_1})-(\ref{eq:demand_case_4}) correspond to Case~1 through Case~4 of Figure~\ref{fig:bundling_cases}, respectively. $q_{1}=\alpha_{11}-\alpha_{21}\exp\left(-\alpha_{31}n_{1}\right)$ and $q_{2}=\alpha_{12}-\alpha_{22}\exp\left(-\alpha_{32}n_{2}\right)$ are the quality of IoT services~1 and 2, respectively.

\begin{prop}
	\label{prop:bundle_optimal}$F_{\mathcal{K}}\left(p_{b},n_{1},n_{2}\right)$ is concave, and hence the closed-form solution of the optimization problem $\underset{p_{b},n_{1},n_{2}}{\text{maximize }}F_{\mathcal{K}}\left(p_{b},n_{1},n_{2}\right)$ is globally optimal.
\end{prop}
\begin{figure*}
	\begin{equation}
	\mathbf{H}_{\mathcal{K}}=\left[\begin{array}{ccc}
	
	\Delta_{1,1} &
	\dfrac{1.5M\alpha_{21}\alpha_{31}p_{b}^2\exp\left(-\alpha_{31}n_1\right)}{q1^2q2} &
	\dfrac{1.5M\alpha_{22}\alpha_{32}p_{b}^2\exp\left(-\alpha_{32}n_2\right)}{q1q2^2} \\
	
	\dfrac{1.5M\alpha_{21}\alpha_{31}p_{b}^2\exp\left(-\alpha_{31}n_1\right)}{q1^2q2} &
	\Delta_{1,2}&
	\dfrac{-0.5M\alpha_{21}\alpha_{22}\alpha_{31}\alpha_{32}p_{b}^3}{q1^2q2^2\exp\left(\alpha_{31}n_1\right)\exp\left(\alpha_{32}n_2\right)}\\
	
	\dfrac{1.5M\alpha_{22}\alpha_{32}p_{b}^2\exp\left(-\alpha_{32}n_2\right)}{q1q2^2} &
	\dfrac{-0.5M\alpha_{21}\alpha_{22}\alpha_{31}\alpha_{32}p_{b}^3}{q1^2q2^2\exp\left(\alpha_{31}n_1\right)\exp\left(\alpha_{32}n_2\right)} &
	\Delta_{1,3}
	
	\end{array}\right].
	\label{eq:hessian_bundle}
	\end{equation}
	\noindent\rule{\textwidth}{1pt}
\end{figure*}

\begin{IEEEproof}
	We will next prove the concavity of the profit function for Case~1 ($C1:p_{b}\leq q_{1}$ and $C2:p_{b}\leq q_{2}$) using Sylvester's criterion which provides a necessary and sufficient condition for the concavity of $F_{\mathcal{K}}\left(\cdot\right)$~\cite{chong2013introduction}. The other demand cases (Cases~2-4) can be analyzed similarly and are omitted due to the space limit. The Hessian matrix~$\mathbf{H}_{\mathcal{K}}$ of $F_{\mathcal{K}}\left(\cdot\right)$ for Case~1 is defined as in (\ref{eq:hessian_bundle}) shown at the top of the next page. $\Delta_{1,1}$ $\Delta_{1,2}$, and $\Delta_{1,3}$ are the three first-order principal minors of $\mathbf{H}_{\mathcal{K}}$ which can be derived as follows:
	\begin{equation}
	\Delta_{1,1} = -\frac{3Mp_{b}}{q_{1}q_{2}}\leq0,
	\end{equation}
	\begin{multline}
	\Delta_{1,2} = 	-\frac{M\alpha_{21}^{2}\alpha_{31}^{2}p_{b}^{3}\exp\left(-2\alpha_{31}n_{1}\right)}{q_{1}^{3}q_{2}} \\
	-\frac{0.5M\alpha_{21}\alpha_{31}^{2}p_{b}^{3}\exp\left(-\alpha_{31}n_{1}\right)}{q_{1}^{2}q_{2}}\leq0,
	\end{multline}
	\begin{multline}
	\Delta_{1,3}= -\frac{M\alpha_{22}^{2}\alpha_{32}^{2}p_{b}^{3}\exp\left(-2\alpha_{32}n_{2}\right)}{q_{1}q_{2}^{3}} \\
	-\frac{0.5M\alpha_{22}\alpha_{32}^{2}p_{b}^{3}\exp\left(-\alpha_{32}n_{2}\right)}{q_{1}q_{2}{}^{2}}\leq0.
	\end{multline}
	There are also three second-order principal minors for $\mathbf{H}_{\mathcal{K}}$ which are obtained as follows:
	\begin{eqnarray}
	\Delta_{2,1}&=&\frac{0.75M^2\alpha_{21}\alpha_{31}^2p_{b}^4\exp\left(-\alpha_{31}n_1\right)\left(\alpha_{11} + q_{1}\right)}{q_{1}^4q_{2}^2}\geq0, \\
	\Delta_{2,2}&=&\frac{0.75M^2\alpha_{22}\alpha_{32}^2p_{b}^4\exp\left(-\alpha_{32}n_2\right)\left(\alpha_{12} + q_{2}\right)}{q_{1}^2q_{2}^4}\geq0, \\
	\Delta_{2,3}&=&\frac{0.25M^2\alpha_{21}\alpha_{22}\alpha_{31}^2\alpha_{32}^2p_{b}^6A_{1}}{q_{1}^{4}q_{2}^{4}}\geq0,
	\end{eqnarray}
	where
	\begin{multline}
	A_{1}=\exp\left(\alpha_{31}n_1 + \alpha_{32}n_2\right)\bigg(\alpha_{11}\alpha_{12}+  \\
	\alpha_{11}\alpha_{22}\exp\left(-\alpha_{32}n_2\right) + 
	\alpha_{12}\alpha_{21}\exp\left(-\alpha_{31}n_1\right)\bigg).
	\end{multline}
	Finally, $\mathbf{H}_{\mathcal{K}}$ has one third-order principle minor defined as follows:
	\begin{eqnarray}
	\Delta_{3,1}& = & \frac{-0.375M^3\alpha_{21}\alpha_{22}\alpha_{31}^2\alpha_{32}^2p_{b}^7A_{2}}
	{q_{1}^{5}q_{2}^{5}}\leq0,
	\end{eqnarray}
	where
	\begin{equation}
	A_{2}=\exp\left(-\alpha_{31}n_1 - \alpha_{32}n_2\right)\left(\alpha_{11}q_{2} + \alpha_{12}q_{1}\right).
	\end{equation}
	
	Accordingly, Sylvester's criterion is satisfied. This proves that $\mathbf{H}_{\mathcal{K}}$ is negative semidefinite and $F_{\mathcal{K}}\left(\cdot\right)$ is concave. Then, the solution of the optimization problem $\underset{p_{b},n_{1},n_{2}}{\text{maximize }}F_{\mathcal{K}}\left(\cdot\right)$ is globally optimal.
\end{IEEEproof}
We next provide the closed-form solutions for the profit maximization problem of bundled services by applying the KKT conditions.

\subsection{Optimal Subscription Fee and Requested Data Sizes}

\subsubsection{Case 1 $\bigg(C1:p_{b}\leq q_{1}$ and $C2:p_{b}\leq q_{2}\bigg)$}

Substituting the customer demand of (\ref{eq:demand_case_1}) in (\ref{eq:bundling_problem}), the Lagrangian dual function is defined as follows:
\begin{multline}
\mathcal{L}_{\mathcal{K}}\left(\left\{ p_{b},n_{1},n_{2}\right\} ,\lambda_{1},\ldots,\lambda_{5}\right)=
-Mp_{b}\left(1-\frac{0.5p_{b}^{2}}{q_{1}q_{2}}\right)+\\n_{1}c_{1}+n_{2}c_{2}+\lambda_{1}\left(p_{b}-q_{1}\right)
+\lambda_{2}\left(p_{b}-q_{2}\right)-\lambda_{3}p_{b}-\lambda_{4}n_{1}-\lambda_{5}n_{2},\label{eq:bundle_case_1}
\end{multline}
where $\lambda_{1},\ldots,\lambda_{5}$ are the Lagrange multipliers of the constraints $C_{1},\ldots,C_{5}$ in (\ref{eq:bundling_problem}). The first derivatives of (\ref{eq:bundle_case_1}) with respect to $p_{b}$, $n_{1}$, and $n_{2}$ are found as follows:

\begin{multline}
\frac{\partial\mathcal{L}_{\mathcal{K}}\left(\cdot\right)}{\partial p_{b}}=\frac{Mp_{b}}{q_{1}q_{2}}-\lambda_{2}-\lambda_{3}-\lambda_{1}\\
-\frac{Mq_{2}\left(-0.5p_{b}+\left(\alpha_{11}-\alpha_{21}\exp\left(-\alpha_{31}n_{1}\right)\right)\right)}{q_{1}q_{2}},\label{eq:der_bundle_case_1_p12}
\end{multline}
\begin{multline}
\frac{\partial\mathcal{L}_{\mathcal{K}}\left(\cdot\right)}{\partial n_{1}}=c_{1}-\lambda_{4}+\alpha_{21}\alpha_{31}\lambda_{1}\exp\left(-\alpha_{31}n_{1}\right)\\
-\frac{M\alpha_{21}\alpha_{31}p_{b}\exp\left(-\alpha_{31}n_{1}\right)}{q_{1}}\\
+\frac{M\alpha_{21}\alpha_{31}p_{b}\exp\left(-\alpha_{31}n_{1}\right)\left(-0.5p_{b}+q_{2}q_{1}\right)}{q_{1}q_{2}},\label{eq:der_bundle_case_1_n1}
\end{multline}
\begin{multline}
\frac{\partial\mathcal{L}_{\mathcal{K}}\left(\cdot\right)}{\partial n_{2}}=c_{2}-\lambda_{5}+\alpha_{22}\alpha_{32}\lambda_{2}\exp\left(-\alpha_{32}n_{2}\right)\\
-\frac{M\alpha_{22}\alpha_{32}p_{b}\exp\left(-\alpha_{32}n_{2}\right)}{q_{2}}\\
+\frac{M\alpha_{22}\alpha_{32}p_{b}\exp\left(-\alpha_{32}n_{2}\right)\left(-0.5p_{b}+q_{1}q_{2}\right)}{q_{1}\left(q_{2}\right)^{2}}.\label{eq:eq:der_bundle_case_1_n2}
\end{multline}
Applying the KKT necessary and sufficient conditions with no active constraints ($\lambda_{1},\ldots,\lambda_{5}=0$) by setting the derivatives in (\ref{eq:der_bundle_case_1_p12})-(\ref{eq:eq:der_bundle_case_1_n2}) to zero, the closed-form solution can be deduced as follows:

\begin{flalign}
n_{1}^{*} & =\frac{1}{\alpha_{31}}\log\left(\frac{\alpha_{21}}{\alpha_{11}}-\frac{\frac{1}{6}\alpha_{21}A_{3}}{\alpha_{11}\alpha_{32}c_{1}}\right),\label{eq:bundling_sol_n1}\\
n_{2}^{*} & =\frac{1}{\alpha_{32}}\log\left(\frac{\alpha_{22}}{\alpha_{12}}-\frac{\frac{1}{6}\alpha_{22}A_{3}}{\alpha_{12}\alpha_{31}c_{2}}\right),\label{eq:bundling_sol_n2}\\
p_{b}^{*} & =-\frac{0.5A_{3}}{m\alpha_{31}\alpha_{32}},\label{eq:bundling_sol_p12}\\
& \lambda_{1},\ldots,\lambda_{5}=0,\label{eq:bundling_sol_lagrange}
\end{flalign}
where
\begin{multline}
A_{3}=3\alpha_{31}c_{2}+3\alpha_{32}c_{1}-\bigg(\frac{8}{2}\alpha_{11}\alpha_{12}M^{2}\alpha_{31}^{2}\alpha_{32}^{2}\\
+9\alpha_{31}^{2}c_{2}^{2}-18\alpha_{31}\alpha_{32}c_{1}c_{2}+9\alpha_{32}^{2}c_{1}^{2}\bigg)^{0.5}.
\end{multline}
The global solution is achieved (Proposition~\ref{prop:bundle_optimal}), and these closed-form expressions satisfy all constraints $C1,\ldots,C5$ of the optimization problem in~(\ref{eq:bundling_problem}).

\subsubsection{Case 2 $\bigg(C1:p_{b}\leq q_{1}$ and $C2:p_{b}\geq q_{2}\bigg)$}

When the customer demand is defined as in (\ref{eq:demand_case_2}) and the optimization constraints are $C1:p_{b}\leq q_{1}$ and $C2:p_{b}\geq q_{2}$, the Lagrangian dual problem of (\ref{eq:bundling_problem}) is defined as follows:

\begin{multline}
\mathcal{L}_{\mathcal{K}}\left(\left\{ p_{b},n_{1},n_{2}\right\} ,\lambda_{1},\ldots,\lambda_{5}\right)=\\
-0.5Mp_{b}\left(1-\frac{p_{b}}{q_{1}}+1-\frac{\left(p_{b}-q_{2}\right)}{q_{1}}\right)+n_{1}c_{1}+n_{2}c_{2}\\
+\lambda_{1}\left(p_{b}-q_{1}\right)-\lambda_{2}\left(p_{b}-q_{2}\right)-\lambda_{3}p_{b}-\lambda{}_{4}n-\lambda_{5}n_{2},\label{eq:bundle_case_2}
\end{multline}
where $\lambda_{1},\ldots,\lambda_{5}$ are the Lagrange multipliers. Taking the derivatives of (\ref{eq:bundle_case_2}) with respect to $p_{b}$, $n_{1}$, and $n_{2}$, the closed-form solution can be found by solving the resulting derivatives as follows: 

\begin{flalign}
n_{1}^{*} & =\frac{1}{\alpha_{31}}\log\left(\frac{0.25\alpha_{21}A_{4}}{\alpha_{32}c_{1}\left(\alpha_{11}-\alpha_{12}\right)}\right),\\
n_{2}^{*} & =\frac{1}{\alpha_{32}}\log\left(\frac{0.5}{c_{2}}\left(M\alpha_{22}\alpha_{32}-\frac{0.5\alpha_{22}A_{4}}{\alpha_{31}\left(\alpha_{11}-\alpha_{12}\right)}\right)\right),\\
p_{b}^{*} & =\frac{0.5A_{4}}{M\alpha_{31}\alpha_{32}}-\frac{2\alpha_{31}c_{2}+2\alpha_{32}c_{1}-M\alpha_{12}\alpha_{31}\alpha_{32}}{M\alpha_{31}\alpha_{32}},\\
\lambda_{1}^{*} & =\frac{0.25A_{4}}{\alpha_{31}\alpha_{32}\left(\alpha_{11}-\alpha_{12}\right)}-0.5M,\\
\lambda_{2}^{*} & =\frac{0.25A_{4}}{\alpha_{31}\alpha_{32}\left(\alpha_{11}-\alpha_{12}\right)},\\
\lambda_{3}^{*} & =\lambda_{4}^{*}=\lambda_{5}^{*}=0,
\end{flalign}
where

\begin{multline}
A_{4}=2\alpha_{31}c_{2}+2\alpha_{32}c_{1}+M\alpha_{11}\alpha_{31}\alpha_{32}-M\alpha_{12}\alpha_{31}\alpha_{32}\\
-2\bigg(0.25M^{2}\alpha_{11}^{2}\alpha_{31}^{2}\alpha_{32}^{2}-0.5M^{2}\alpha_{11}\alpha_{12}\alpha_{31}^{2}\alpha_{32}^{2}\\
+0.25M^{2}\alpha_{12}^{2}\alpha_{31}^{2}\alpha_{32}^{2}+M\alpha_{11}\alpha_{31}^{2}\alpha_{32}c_{2}-M\alpha_{11}\alpha_{31}\alpha_{32}^{2}c_{1}\\
-M\alpha_{12}\alpha_{31}^{2}\alpha_{32}c_{2}+M\alpha_{12}\alpha_{31}\alpha_{32}^{2}c_{1}+\alpha_{31}^{2}c_{2}^{2}\\
+2\alpha_{31}\alpha_{32}c_{1}c_{2}+\alpha_{32}^{2}c_{1}^{2}\bigg)^{0.5}.
\end{multline}

\subsubsection{Case 3 $\bigg(p_{b}\geq q_{1}$ and $p_{b}\leq q_{2}\bigg)$}

In this case, the Lagrangian dual function of (\ref{eq:bundling_problem}) and (\ref{eq:demand_case_3}) can be written as:

\begin{multline}
\mathcal{L}_{\mathcal{K}}\left(\left\{ p_{b},n_{1},n_{2}\right\} ,\lambda_{1},\ldots,\lambda_{5}\right)=\\
-0.5Mp_{b}\left(1-\frac{p_{b}-q_{1}}{q_{2}}+1-\frac{p_{b}}{q_{2}}\right)+n_{1}c_{1}+n_{2}c_{2}\\
-\lambda_{1}\left(p_{b}-q_{1}\right)+\lambda_{2}\left(p_{b}-q_{2}\right)-\lambda_{3}p_{b}-\lambda{}_{4}n-\lambda_{5}n_{2}.\label{eq:bundle_case_3}
\end{multline}
Taking the derivatives of (\ref{eq:bundle_case_3}) with respect to $p_{b}$, $n_{1}$, and $n_{2}$, the closed-form solution can be expressed as follows: 
\begin{flalign}
n_{1}^{*} & =\frac{1}{\alpha_{31}}\log\left(\frac{0.25\alpha_{21}A_{5}}{\alpha_{32}c_{1}\left(\alpha_{11}-\alpha_{12}\right)}\right),\\
n_{2}^{*} & =\frac{1}{\alpha_{32}}\log\left(\frac{0.5}{c_{2}}\left(M\alpha_{22}\alpha_{32}-\frac{0.5\alpha_{22}A_{5}}{\left(\alpha_{31}\left(\alpha_{11}-\alpha_{12}\right)\right)}\right)\right),\\
p_{b}^{*} & =\frac{0.5A_{5}}{M\alpha_{31}\alpha_{32}}-\frac{2\alpha_{31}c_{2}+2\alpha_{32}c_{1}-M\alpha_{12}\alpha_{31}\alpha_{32}}{M\alpha_{31}\alpha_{32}},\\
\lambda_{1}^{*} & =0.5M-\frac{0.25A_{5}}{\alpha_{31}\alpha_{32}\left(\alpha_{11}-\alpha_{12}\right)},\\
\lambda_{2}^{*} & =-\frac{0.25A_{5}}{\alpha_{31}\alpha_{32}\left(\alpha_{11}-\alpha_{12}\right)},\\
\lambda_{3}^{*} & =\lambda_{4}^{*}=\lambda_{5}^{*}=0,
\end{flalign}
where
\begin{multline}
A_{5}=2\alpha_{31}c_{2}+2\alpha_{32}c_{1}+M\alpha_{11}\alpha_{31}\alpha_{32}-M\alpha_{12}\alpha_{31}\alpha_{32}\\
-2\bigg(0.25M^{2}\alpha_{11}^{2}\alpha_{31}^{2}\alpha_{32}^{2}-0.5M^{2}\alpha_{11}\alpha_{12}\alpha_{31}^{2}\alpha_{32}^{2}\\
+0.25M^{2}\alpha_{12}^{2}\alpha_{31}^{2}\alpha_{32}^{2}+M\alpha_{11}\alpha_{31}^{2}\alpha_{32}c_{2}-M\alpha_{11}\alpha_{31}\alpha_{32}^{2}c_{1}\\
-M\alpha_{12}\alpha_{31}^{2}\alpha_{32}c_{2}+M\alpha_{12}\alpha_{31}\alpha_{32}^{2}c_{1}+\alpha_{31}^{2}c_{2}^{2}\\
+2\alpha_{31}\alpha_{32}c_{1}c_{2}+\alpha_{32}^{2}c_{1}^{2}\bigg)^{0.5}.
\end{multline}

\subsubsection{Case 4 $\bigg(p_{b}\geq q_{1}$ and $p_{b}\geq q_{2}\bigg)$}

With (\ref{eq:bundling_problem}) and (\ref{eq:demand_case_4}), the Lagrangian dual function can be formulated as follows:

\begin{multline}
\mathcal{L}_{\mathcal{K}}\left(\left\{ p_{b},n_{1},n_{2}\right\} ,\lambda_{1},\ldots,\lambda_{5}\right)=\\
-0.5Mp_{b}\left(1-\frac{\left(p_{b}-q_{1}\right)}{q_{2}}\right)\left(1-\frac{\left(p_{b}-q_{2}\right)}{q_{1}}\right)+n_{1}c_{1}+n_{2}c_{2}\\
-\lambda_{1}\left(p_{b}-q_{1}\right)-\lambda_{2}\left(p_{b}-q_{2}\right)-\lambda_{3}p_{b}-\lambda{}_{4}n-\lambda_{5}n_{2}.\label{eq:bundle_case_4}
\end{multline}
Taking the derivatives of (\ref{eq:bundle_case_4}) with respect to
$p_{b}$, $n_{1}$, and $n_{2}$, the closed-form solution can be
deduced as follows:
\begin{flalign}
n_{1}^{*} & =\frac{1}{\alpha_{31}}\log\left(\frac{0.25\alpha_{21}\alpha_{31}A_{6}}{c_{1}\left(\alpha_{11}\alpha_{31}\alpha_{32}-\alpha_{12}\alpha_{31}\alpha_{32}\right)}\right),\\
n_{2}^{*} & =\frac{1}{\alpha_{32}}\log\left(\frac{0.5}{c_{2}}\left(m\alpha_{22}\alpha_{32}-\frac{0.5\alpha_{22}\alpha_{32}A_{6}}{\alpha_{11}\alpha_{31}\alpha_{32}-\alpha_{12}\alpha_{31}\alpha_{32}}\right)\right),\\
p_{b}^{*} & =\alpha_{11}-\frac{\alpha_{21}c_{1}\left(\alpha_{11}\alpha_{31}\alpha_{32}-1.0\alpha_{12}\alpha_{31}\alpha_{32}\right)}{0.25\alpha_{21}\alpha_{31}A_{6}},\\
\lambda_{1}^{*} & =\frac{0.25A_{6}}{\alpha_{11}\alpha_{31}\alpha_{32}-\alpha_{12}\alpha_{31}\alpha_{32}}-0.5m,\\
\lambda_{2}^{*} & =-\frac{0.25A_{6}}{\alpha_{11}\alpha_{31}\alpha_{32}-\alpha_{12}\alpha_{31}\alpha_{32}},\\
\lambda_{3}^{*} & =\lambda_{4}^{*}=\lambda_{5}^{*}=0,
\end{flalign}
where
\begin{multline}
A_{6}=2\alpha_{31}c_{2}+2\alpha_{32}c_{1}+M\alpha_{11}\alpha_{31}\alpha_{32}-M\alpha_{12}\alpha_{31}\alpha_{32}\\
-\bigg(M^{2}\alpha_{11}^{2}\alpha_{31}^{2}\alpha_{32}^{2}-2M^{2}\alpha_{11}\alpha_{12}\alpha_{31}^{2}\alpha_{32}^{2}+M^{2}\alpha_{12}^{2}\alpha_{31}^{2}\alpha_{32}^{2}\\
+4M\alpha_{11}\alpha_{31}^{2}\alpha_{32}c_{2}-4M\alpha_{11}\alpha_{31}\alpha_{32}^{2}c_{1}-4M\alpha_{12}\alpha_{31}^{2}\alpha_{32}c_{2}\\
+4M\alpha_{12}\alpha_{31}\alpha_{32}^{2}c_{1}+4\alpha_{31}^{2}c_{2}^{2}+8\alpha_{31}\alpha_{32}c_{1}c_{2}+4\alpha_{32}^{2}c_{1}^{2}\bigg)^{0.5}.
\end{multline}

The cooperative providers should run the optimization of the four
cases derived in this section. The case with the maximum resulting
profit should be used in the bundled service.

\section{Profit Sharing Among IoT Providers\label{sec:revenue_sharing }}

After forming a coalition $\mathcal{K}$ to sell IoT services as
a bundle, the cooperative providers share the resulting profit. This
section presents a profit sharing model using the core solution and
Shapley value from cooperative game theory~\cite{myerson2013game}
to define the payoff allocations for the cooperative providers in
$\mathcal{K}$.

\subsection{The Core Solution}

Let $\varphi_{k}$ indicate the profit share of service provider $k\in\mathcal{K}$.
The core solution set is calculated as follows~\cite{myerson2013game}:
\begin{equation}
\mathcal{C}=\bigg\{\mathbf{\varphi}\mid\underbrace{\sum_{k\in\mathcal{K}}\varphi_{k}=F_{\mathcal{K}}^{*}}_{\text{group rationality}}\text{ \ensuremath{\land}\ }\underbrace{\sum_{k\in\mathcal{S}}\varphi_{k}\geq F_{\mathcal{S}}^{*},\mathcal{S}\subseteq\mathcal{K}}_{\text{individual rationality}}\bigg\}\label{eq:core_sol}
\end{equation}
where $\mathbf{\varphi}\in\mathbb{R}^{2}$ is a vector of profit allocations
$\varphi_{k}$, $F_{\mathcal{K}}^{*}$ is the profit of bundling,
and $F_{\mathcal{S}}^{*}$ is the profit of separate selling when
$\left|\mathcal{S}\right|=1$. $\mathcal{C}$ includes a set of profit
allocations which guarantee no service provider will reject the payoff
allocation, i.e., no incentive of leaving the coalition to sell services
separately.

The core solution can be empty, containing a large number of possible
solutions, or unfair to a service provider based on the individual
contributions to the bundle formulation. Therefore, we next present
the Shapley solution which provides a single and fair solution of
the profit sharing problem.

\subsection{The Shapley Solution}

The Shapley solution provides a fair allocation of the bundling profit
among the service providers forming a bundling coalition $\mathcal{K}$.
For each cooperative provider $k\in\mathcal{K}$, the Shapley value
$\eta$ assigns a payoff $\eta_{k}$ found as~\cite{myerson2013game}:
\begin{equation}
\eta_{k}=\sum_{\mathcal{S}\subseteq\mathcal{K}\setminus\left\{ k\right\} }\underbrace{\frac{\left|\mathcal{S}\right|!\left(\left|\mathcal{K}\right|-\left|\mathcal{S}\right|-1\right)!}{\left|\mathcal{K}\right|!}}_{\text{probability of ordering}}\underbrace{\left(F_{\mathcal{S}\cup\left\{ k\right\} }^{*}-F_{\mathcal{S}}^{*}\right)}_{\text{contribution}}.\label{eq:shapley_sol}
\end{equation}
The first term defines the probability of ordering providers to form
the bundle coalition. The second term defines the marginal contribution
of each provider $k\in\mathcal{K}$ to the bundle. The Shapley solution
is efficient such that $\sum_{k\in\mathcal{K}}\eta_{k}=F_{\mathcal{K}}^{*}$.
\section{Numerical Results and Discussion\label{sec:numerical_results}}

\begin{figure*}
	\begin{centering}
		\includegraphics[width=0.82\paperwidth,trim=2cm 1cm 0.5cm 0cm]{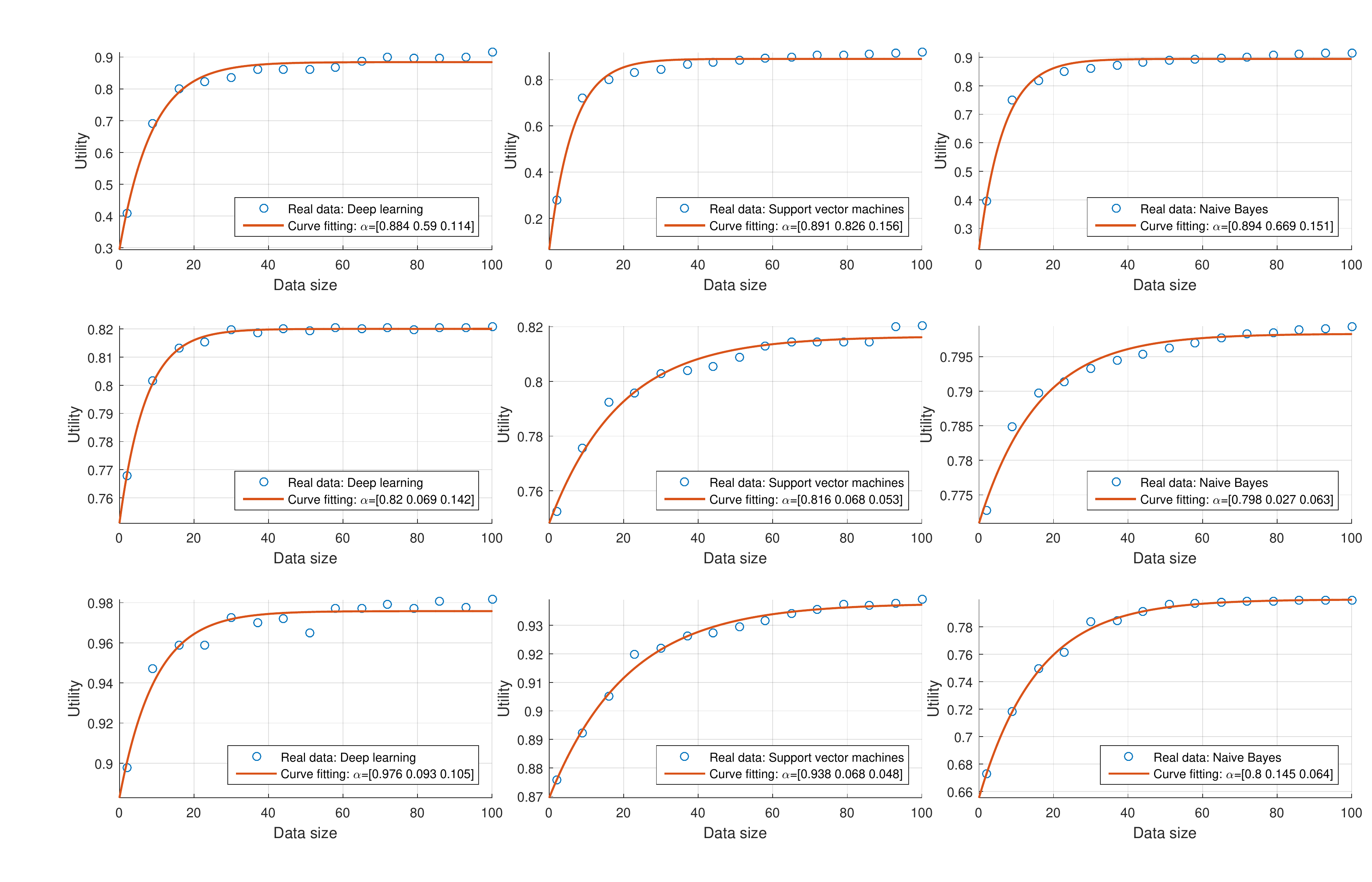}
		\par\end{centering}
	
	\centering{}\caption{Estimating the utility of data $q(n;\mathbf{\alpha})$ of three popular
		classifiers. The first row is for Service~1, the second row is for
		Service~2, and the third row is for Service~3, respectively.\label{fig:data_utility}}
\end{figure*}

This section presents extensive numerical experiments to evaluate the performance of the proposed market model and pricing schemes for selling IoT services separately and as a bundle.

\subsection{Datasets, machine learning, and Market Setup}

We run experiments on three IoT services trained on real-world datasets. Today there are a few standard machine learning models that are widely used across tasks in IoT. To build the IoT service, we use three popular machine learning algorithms, namely, deep learning~\cite{Goodfellow-et-al-2016-Book}, support vector machines~\cite{steinwart2008support}, and multinomial naive Bayes~\cite{kibriya2004multinomial}.
\begin{itemize}
	\item \emph{Service 1 (text categorization)}: IoT requires context awareness and consistent understanding of the user data~\cite{perera2014context}. We build Service~1 that predicts the user context from text data. Service~1 is trained on the 20~newsgroups dataset~\cite{newsgroups20}. The dataset includes $18,828$ data samples classified into $20$ topics, e.g.,~sports, computer hardware, politics, etc. We use $13,180$ samples for training the machine learning models and $5,648$ samples for testing and accuracy prediction. A data unit contains $188$~samples, i.e., one percent of the full dataset samples. We assume that the data vendor offers this dataset with a data unit cost of $c_{1}= 0.1$\footnote{It is in monetary unit which is currency independent.}.
	
	\item \emph{Service 2 (sentiment analysis)}: IoT wearables provide a robust tool for determining the sentiment aspects, e.g.,~opinion, feedback, and emotion, of users~\cite{alhanai2017predicting}. Sentiment predicting IoT has many applications in smart homes, recommender systems, and content-based filtering. We build a data service to predict the sentiment (either positive or negative polarities) by training the machine learning models on the Sentiment140 dataset~\cite{go2009twitter}. The dataset contains $629,146$~tweet samples for model training and $419,431$ samples for testing. Each data unit includes $10,482$~samples. We assume that a data unit has a cost value of $c_{2}=0.05$.
	
	\item \emph{Service~3 (Optical character recognition)}: IoT devices generally contain high-resolution cameras for capturing image and video data. The recent advances in machine learning enable extracting useful information from visual data. Service~3 is an IoT service that extracts text from handwritten images, e.g.,~this is needed for monitoring the expiration date of food in a smart fridge~\cite{luo2008smart}. We train the machine learning models using the MNIST dataset~\cite{lecun2010mnist}. The dataset contains $50,000$ training image samples and $10,000$ testing images. A data unit includes $600$~samples.
\end{itemize}
Text classification (Services~1 and 2) typically requires transforming the text into numerical features suitable for different classifiers. In our experiments, we use a popular feature set known as term frequency\textendash inverse document frequency (TF-IDF) \cite{rajaraman2012mining} which counts the number of times that a word appears in a text. Unless otherwise stated, we assume a base of $M=50$ customers. 

\begin{figure}
	\begin{centering}
		\subfloat[Services~1 (AIC is 397.5)]{\begin{centering}
				\includegraphics[width=0.8\columnwidth,trim=0cm 0.5cm 0cm 1cm]{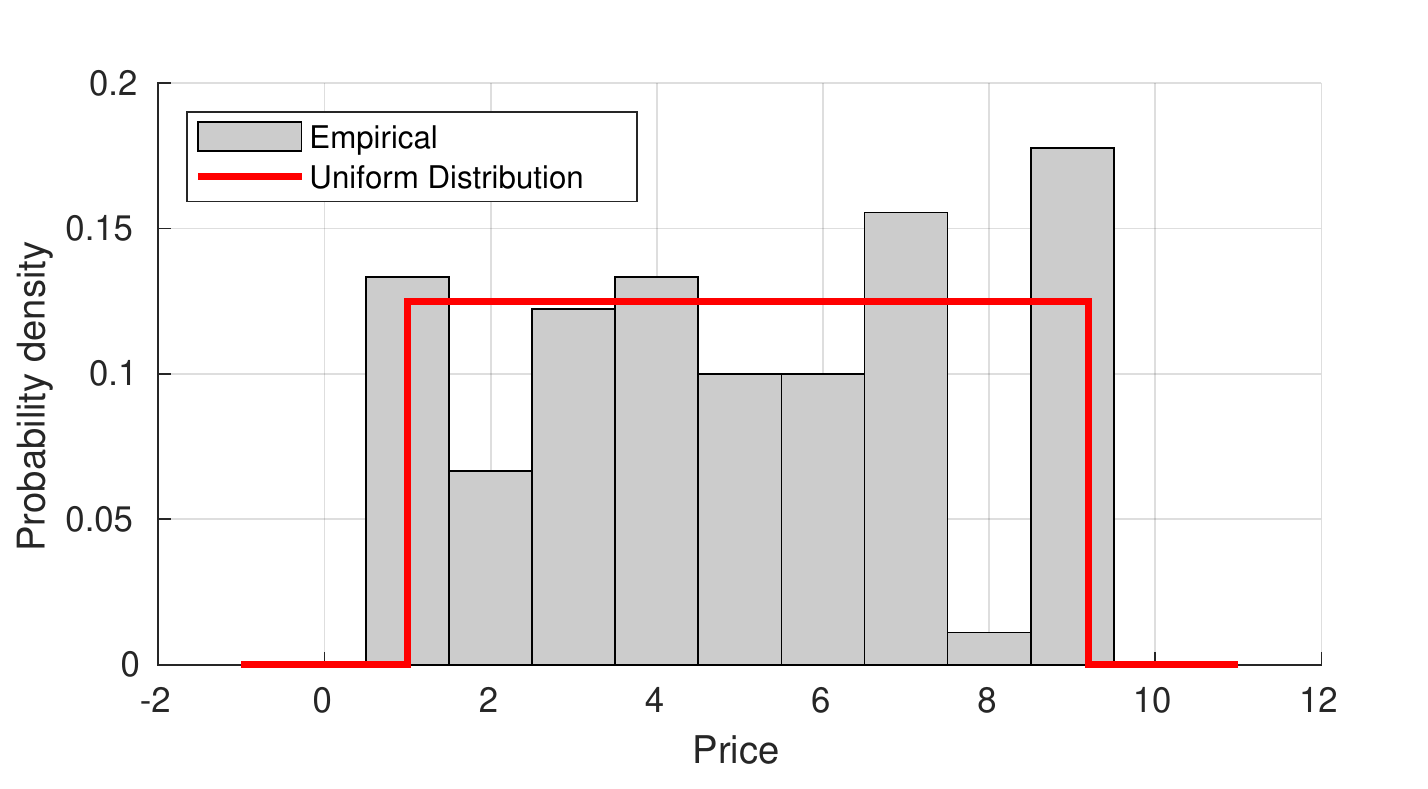}
				\par\end{centering}
		}
	
		\subfloat[Services~2 (AIC is 274.5)]{\begin{centering}
				\includegraphics[width=0.8\columnwidth,trim=0cm 0.5cm 0cm 0.5cm]{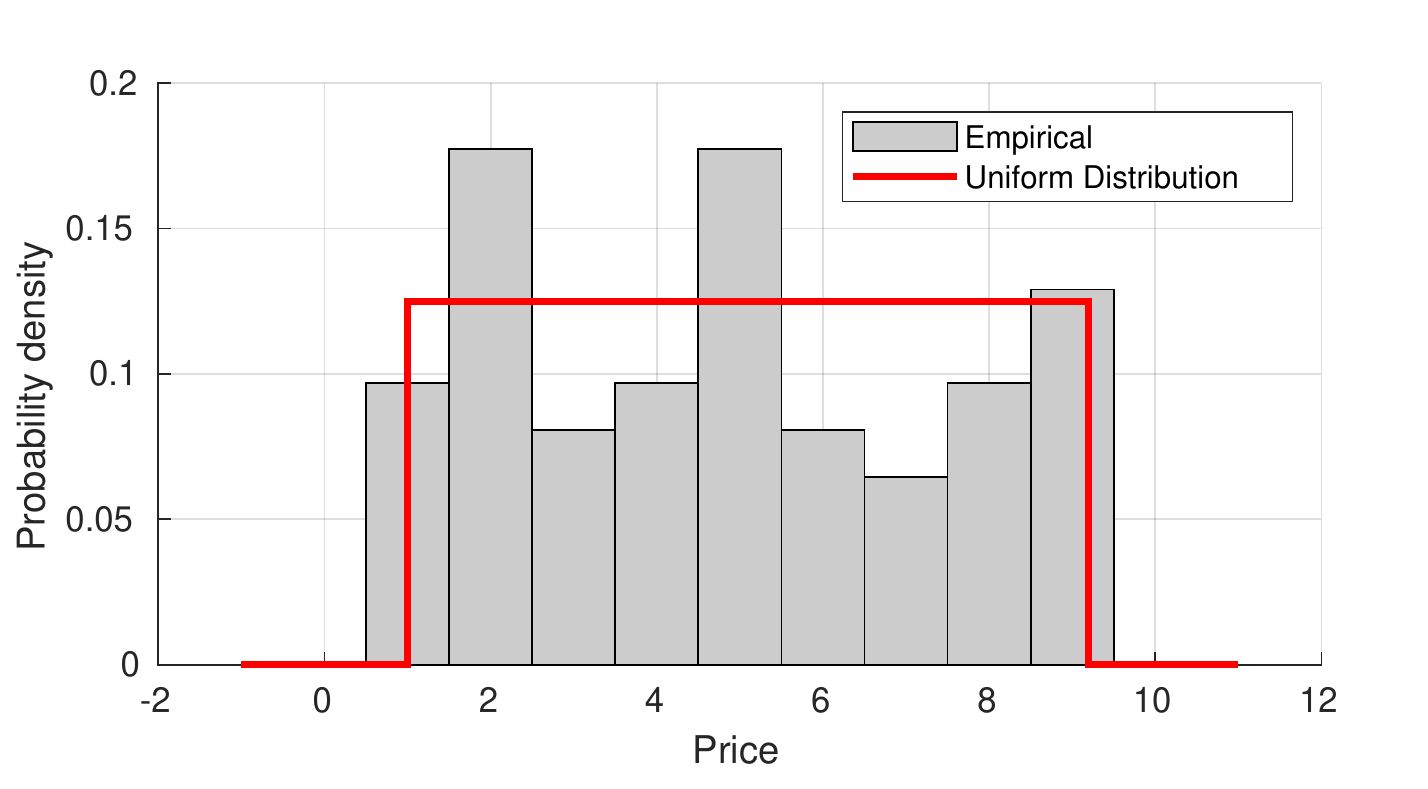}
				\par\end{centering}
		}
		\par\end{centering}
	
	\caption{Modeling the willingness-to-pay distribution based on real-world (empirical) customer surveys. $\theta$ of Services~1 and 2 follows a uniform distribution. \label{fig:wtp_surveys}}
\end{figure}

Figure~\ref{fig:wtp_surveys} shows the willingness-to-pay responses to our real-world customer surveys conducted using Google Surveys\footnote{\url{https://surveys.google.com}} for estimating the willingness-to-pay $\theta$ distribution in IoT. In each survey, we asked 100~users open-ended questions that elicit a direct expression of value. For example, to find $\theta$ of Service~2, we used the question of ``\textit{how much would you be willing to pay for a wearable device that plays music and suggests activities (e.g., sports and readings) based on your sentiment and mood?}''. Figure~\ref{fig:wtp_surveys} shows that $\theta$ follows a uniform distribution. For model selection, we use the Akaike information criterion (AIC)~\cite{akaike1998information}. We tested many relevant data distributions including the negative binomial, poisson, and geometric distributions. The uniform distribution has the minimum AIC values of 397.5 and 274.5 for Services~1 and 2, respectively.

\subsection{Service Quality\label{sub:exp_data_utility}}

Figure~\ref{fig:data_utility} shows our extensive experiments on finding the quality of Services~1-3. The recognition accuracy increases as the requested data size increases and vice versa. When the requested data size is increased, the increase in the accuracy diminishes. Besides, we note that the utility function defined in (\ref{eq:data_quality}) fits the series of real data points well. We also observe a similar diminishing increase of accuracy for other machine learning algorithms such as k-nearest neighbors, logistic regression, random forests, and passive aggressive classifiers. However, due to the space limit, we omit them from this paper.

Based on these results, for the rest of our experiments we use $q_{1}=0.884-0.59\exp\left(-0.114n_{1}\right)$ and $q_{2}=0.82-0.069\exp\left(-0.142n_{2}\right)$ for service qualities which correspond to using deep learning on the 20~newsgroups (Service~1) and Sentiment140 (Service~2) datasets.

\subsection{Separate Selling of Services}

In this section, we study the separate sales of Service~1 under the model and pricing schemes of Section~\ref{sec:separate_selling}. Assuming a deep learning implementation of Service~1, the service quality parameters are $\alpha_{1}=0.884$, $\alpha_{2}=0.59$ , and $\alpha_{3}=0.114$ as specified in Figure~\ref{fig:data_utility}.

\subsubsection{Profit Maximization}

\begin{figure}
	\begin{centering}
		\includegraphics[width=1.0\columnwidth,trim=0cm 0cm 0cm 0cm]{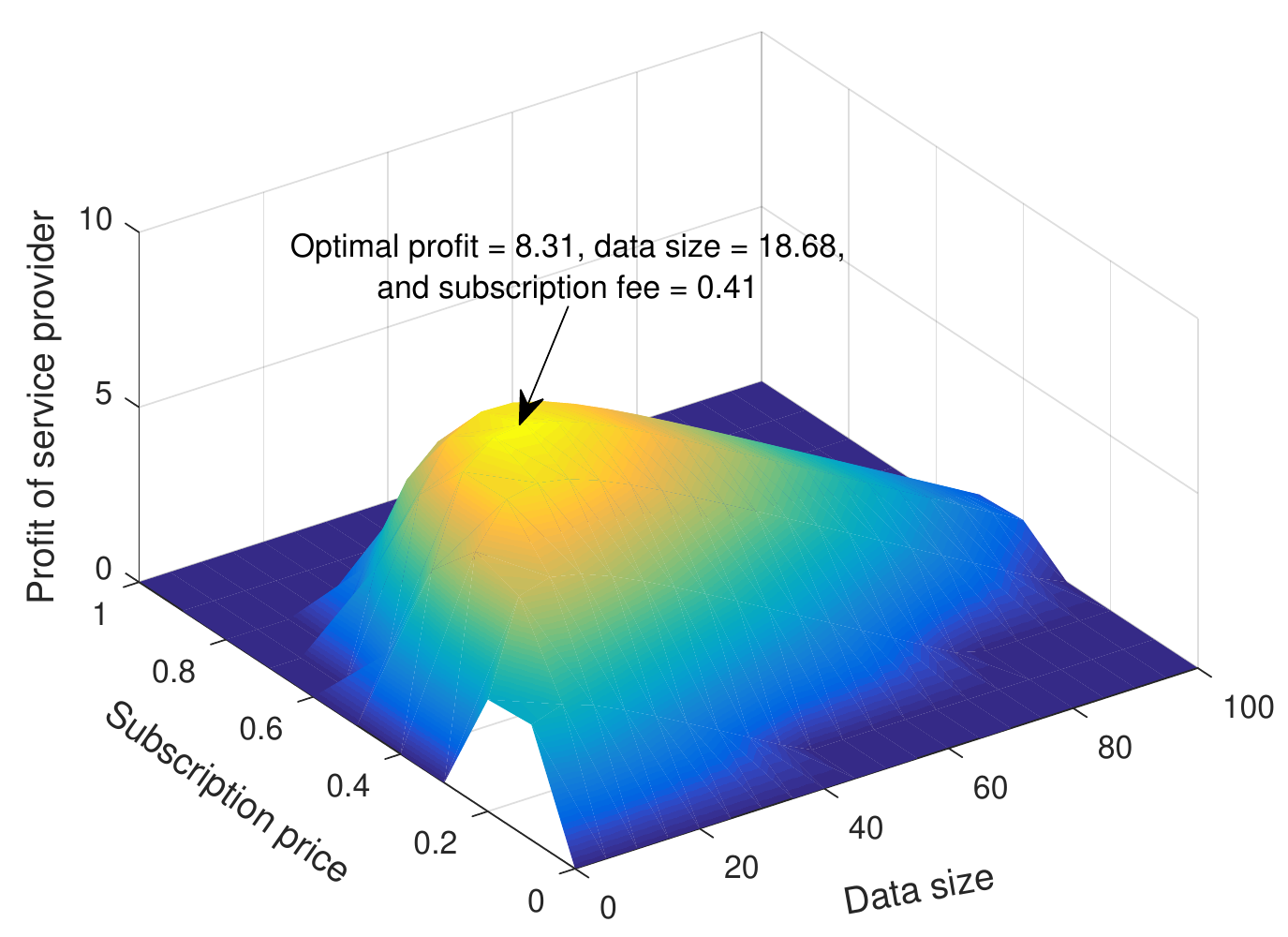}
		\par\end{centering}
	
	\caption{Profit $F\left(p_{s},n\right)$ of a monopolist service provider under varied data sizes and subscription fees.\label{fig:seperate_profit_fee}}

\end{figure}

In Figure~\ref{fig:seperate_profit_fee}, we analyze the profit $F\left(p_{s},n\right)$ of Service~1 under varied requested data sizes $n$ and subscription fees $p_{s}$. When the subscription fee is high, fewer customers will be interested in buying the service which will accordingly decrease the resulting profit. A low subscription fee has a similar negative effect on the profit even though more customers will decide to buy the service. Likewise, a poor spending on buying data from the vendor results in poor profit for the provider as fewer customers will buy the service due to its low accuracy. When the requested data size is high, the service accuracy will increase, but the profit will be negatively affected due to the high data cost. Recall that the profit is the revenue value that remains after accounting for the data cost as in (\ref{eq:seperate_profit}).

Using (\ref{eq:seperate_sol_n}) and (\ref{eq:eq:seperate_sol_ps}) to determine the optimal service setting, we find that the optimal data size $n^{*}=18.68$ and subscription fee $p_{s}^{*}=0.41$ resulting in an optimal profit of $F\left(n^{*},p_{s}^{*}\right)=8.31$.

As data is the main commodity in IoT services, we next study the impact of data price (Section~\ref{sub:impact_data_cost}) and data quality (Section~\ref{sub:impact_data_quality}) in the economics of a monopolist service provider.

\subsubsection{The Impact of Data Cost\label{sub:impact_data_cost}}

\begin{figure}
	\begin{centering}
		\includegraphics[width=0.9\columnwidth,trim=0.5cm 0.5cm 0.5cm 0cm]{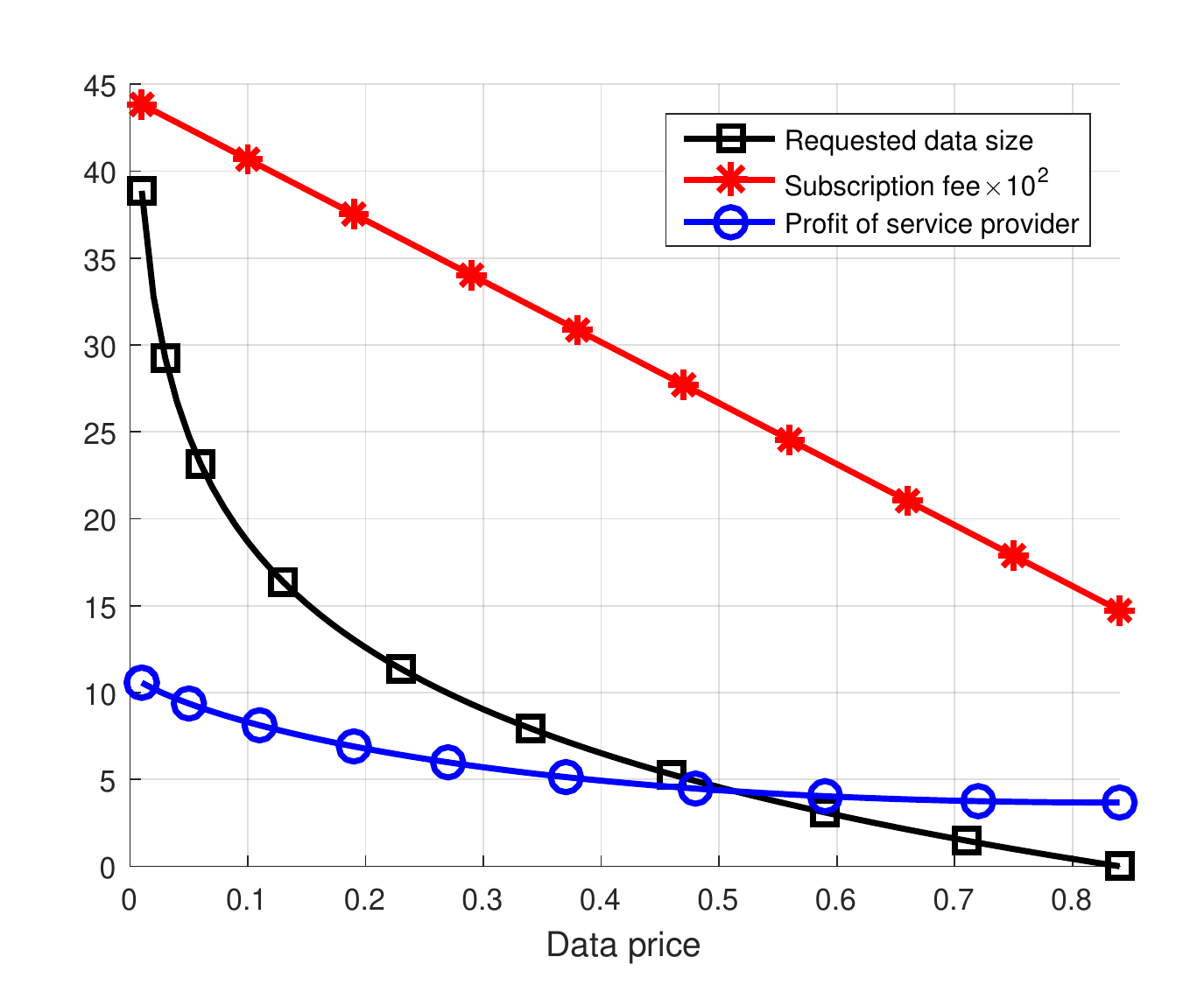}
		\par\end{centering}
	
	\caption{Optimal profit $F\left(p_{s}^{*},n^{*}\right)$ of the service provider, subscription fee $p_{s}^{*}$, and requested data size $n^{*}$ under varied data prices.\label{fig:seperate_data_price}}
\end{figure}

Figure~\ref{fig:seperate_data_price} shows the impact of the data cost $c$ on the optimal solutions, i.e., optimal requested data size $n^{*}$ and subscription fee $p_{s}^{*}$, and the resulting profit $F\left(n^{*},p_{s}^{*}\right)$ from the separate sales of Service~1. Firstly, we observe that the service provider will make a lower profit out of the service if the data price is increased. Secondly, when the data price is high, the service provider will try to decrease the cost by requesting less data from the data vendor. Then, as requesting less data decreases the service quality $q$, the service provider should also decrease the subscription fee in order to attract more customer subscriptions. This is intuitive as the customers infer both the subscription fee and the service quality when making their subscription decision as defined in (\ref{eq:seperate_profit}). Increasing the data price beyond $c=0.84$ produces infeasible solutions, i.e., negative profit, as the constraint $M\alpha_{2}\alpha_{3}\geq4c$ will not be satisfied.

\begin{figure}
	\begin{centering}
		\includegraphics[width=0.9\columnwidth,trim=0.5cm 0.5cm 0.5cm 0cm]{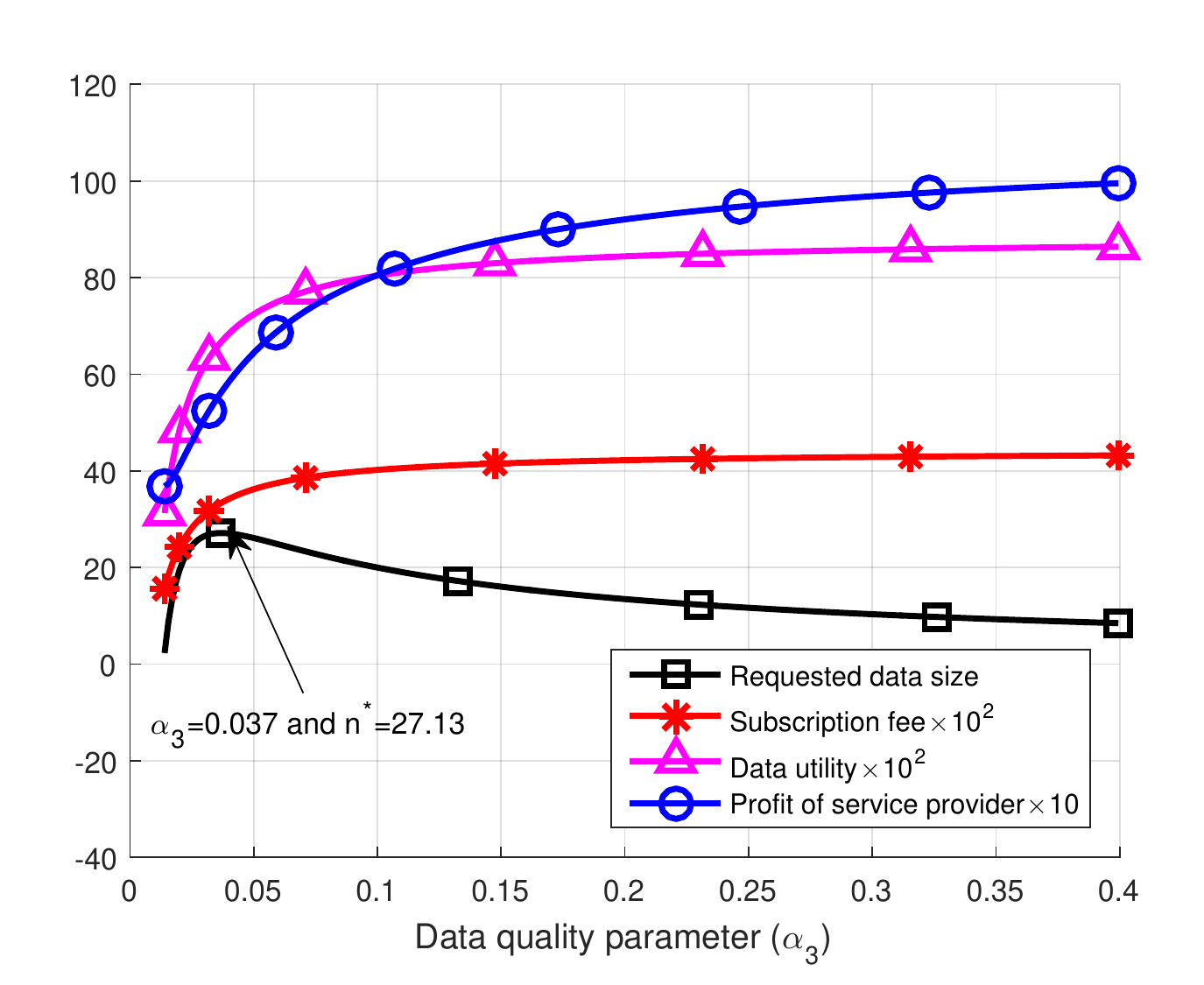}
		\par\end{centering}
	
	\caption{Optimal profit $F\left(p_{s}^{*},n^{*}\right)$ of the service provider, subscription fee $p_{s}^{*}$, requested data size $n^{*}$, and data utility $q\left(n^{*},\alpha\right)$ under varied data quality parameter $\alpha_{3}$.\label{fig:seperate_utility_param}}
\end{figure}

\subsubsection{The Impact of Data Quality\label{sub:impact_data_quality}}

We consider the impact of buying data of different qualities $q$ by changing the data quality parameter $\alpha_{3}$. Specifically, higher values of $\alpha_{3}$ result in higher data qualities $q$ as specified in (\ref{eq:data_quality}). We observe two important results from Figure~\ref{fig:seperate_utility_param}. Firstly, the service provider should always look for high quality data as it helps them to maximize their profit as less data is required for service fitting. If the provider buys more quality data up to a level ($n^{*}=27.13$ at $\alpha_{3}=0.037$), then less data will be requested as enhancing the service quality beyond this level will not be essential for most customers. This is because the customers infer both the service quality and the subscription fee. Secondly, when the data quality is high, less data is required to achieve a satisfactory quality. The low spending in buying data maximizes the provider's profit.

\subsubsection{The Impact of Customer Base Size}

\begin{figure}
	\begin{centering}
		\includegraphics[width=1.0\columnwidth,trim=0cm 0cm 0cm 0cm]{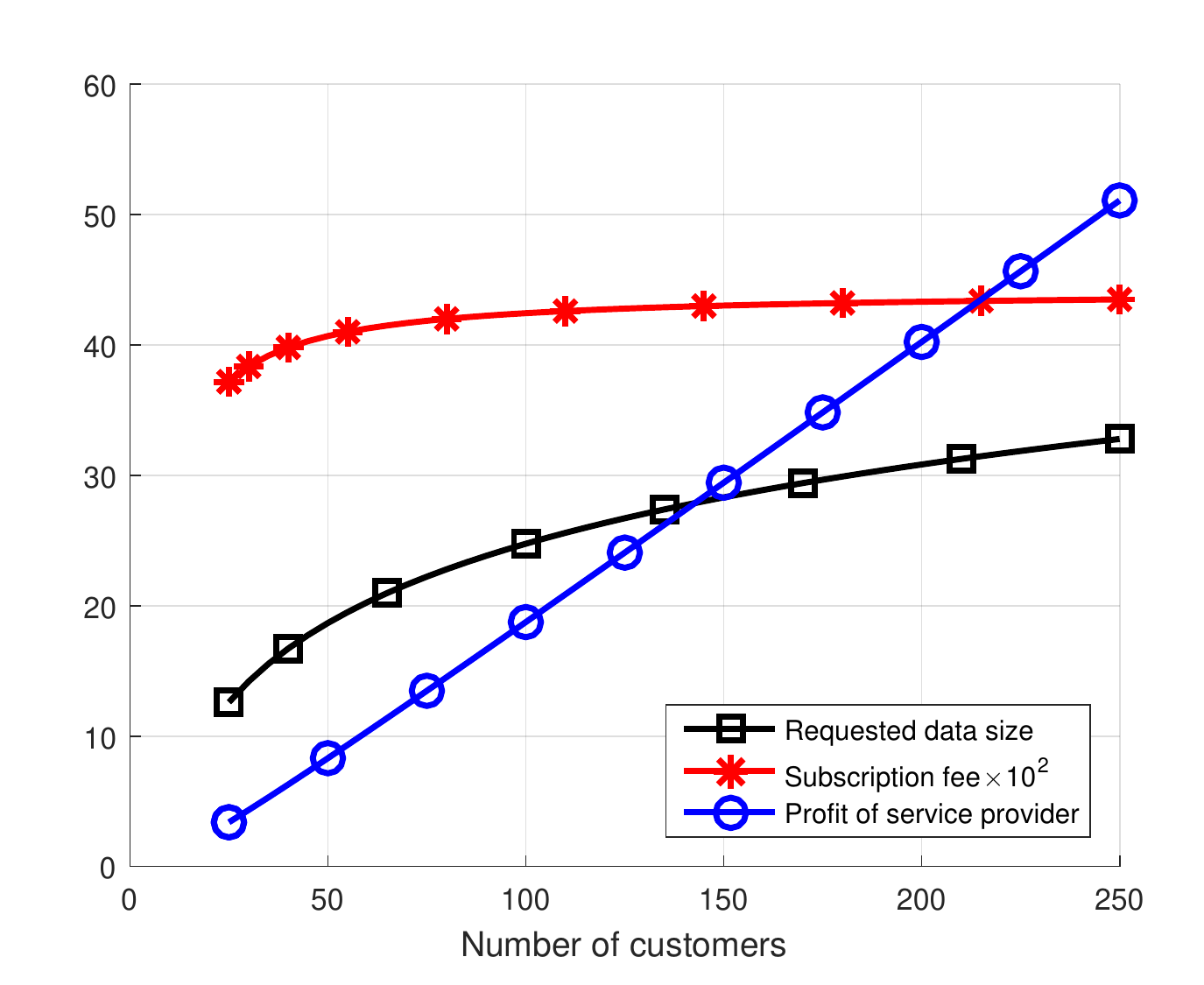}
		\par\end{centering}
	
	\caption{Optimal profit $F\left(p_{s}^{*},n^{*}\right)$ of the service provider, subscription fee $p_{s}^{*}$, and requested data size $n^{*}$ under varied customer bases $M$.\label{fig:seperate_customers}}
\end{figure}

Figure~\ref{fig:seperate_customers} shows the optimal solutions, i.e., optimal requested data size $n^{*}$ and subscription fee $p_{s}^{*}$, and the generated profit $F\left(n^{*},p_{s}^{*}\right)$ under varied sizes of the customer base $M$. We observe that the service provider will increase his investment in buying more data as the customer size increases (more customer demand). Likewise, the subscription fee $p_{s}^{*}$ of the service will be increased as a result of the increased demand. This increase in demand positively impacts the resulting profit of the service provider.

\begin{figure*}
	\begin{centering}
		\subfloat[]{\begin{centering}
				\includegraphics[width=0.95\columnwidth,trim=0cm 0.5cm 0cm 1cm]{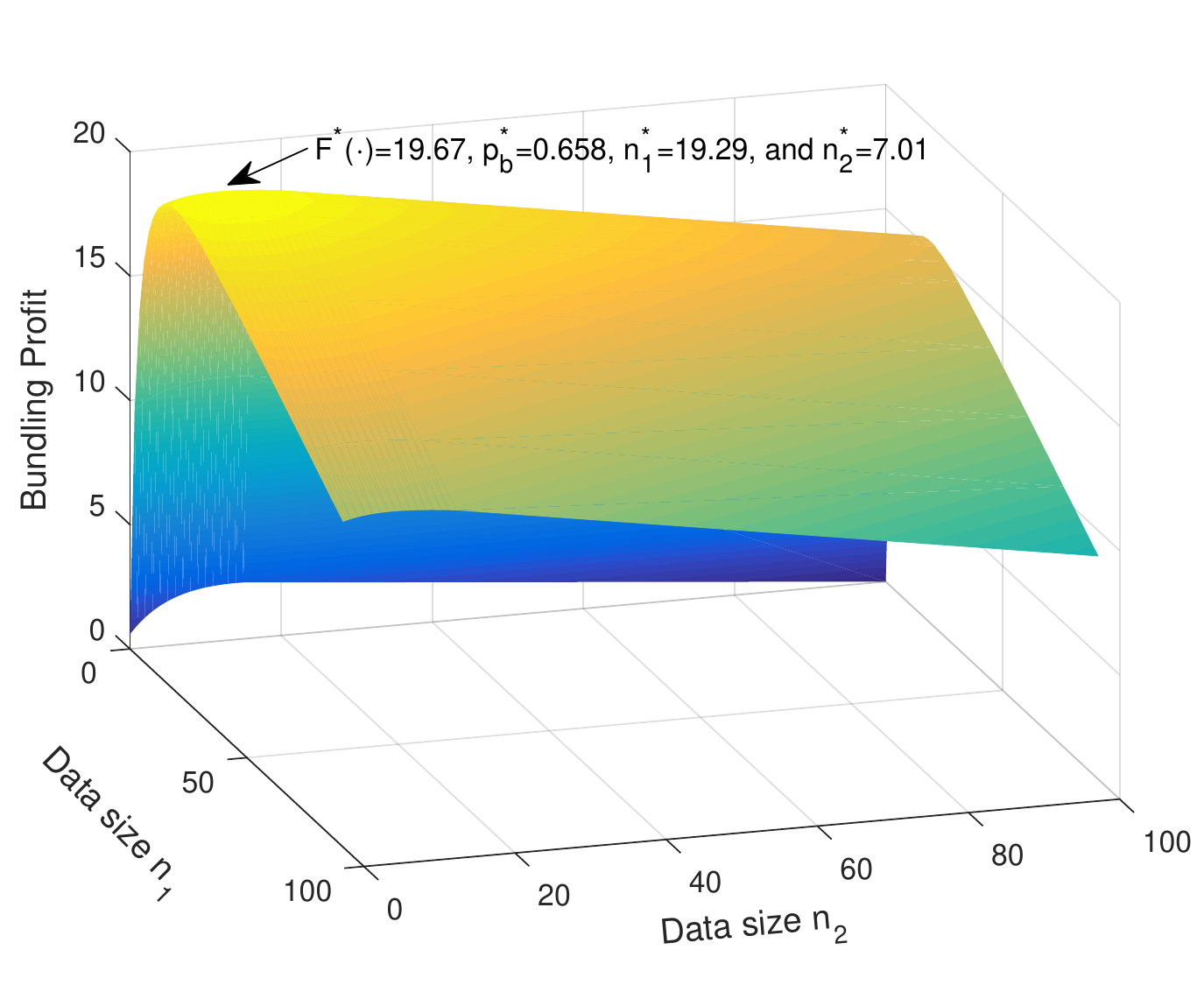}
				\par\end{centering}
		}
		\subfloat[]{\begin{centering}
				\includegraphics[width=0.95\columnwidth,trim=0cm 0.5cm 0cm 0.5cm]{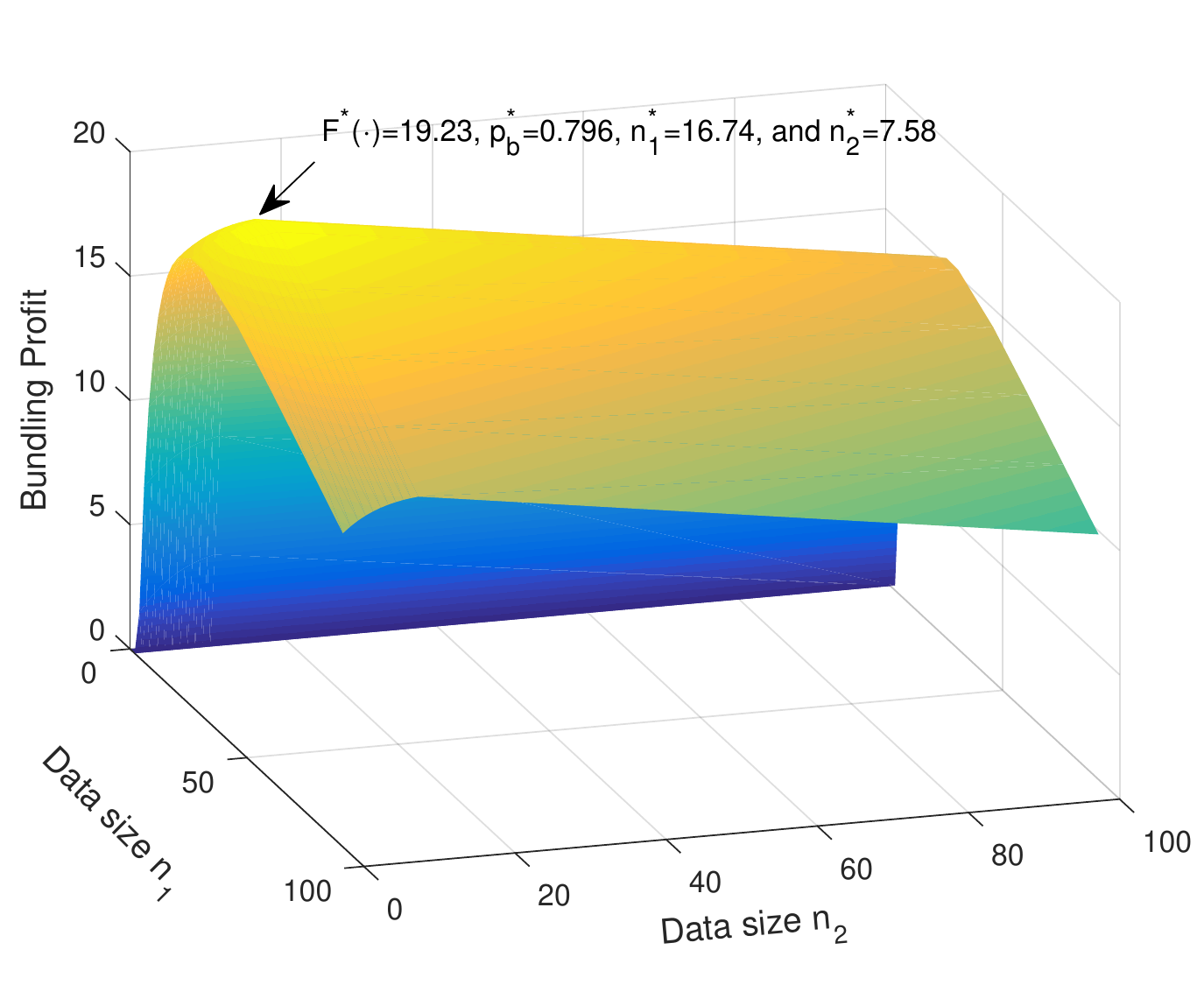}
				\par\end{centering}
		}
		\par\end{centering}
	
	\begin{centering}
		\subfloat[]{\begin{centering}
				\includegraphics[width=0.95\columnwidth,trim=0cm 0.5cm 0cm 0.5cm]{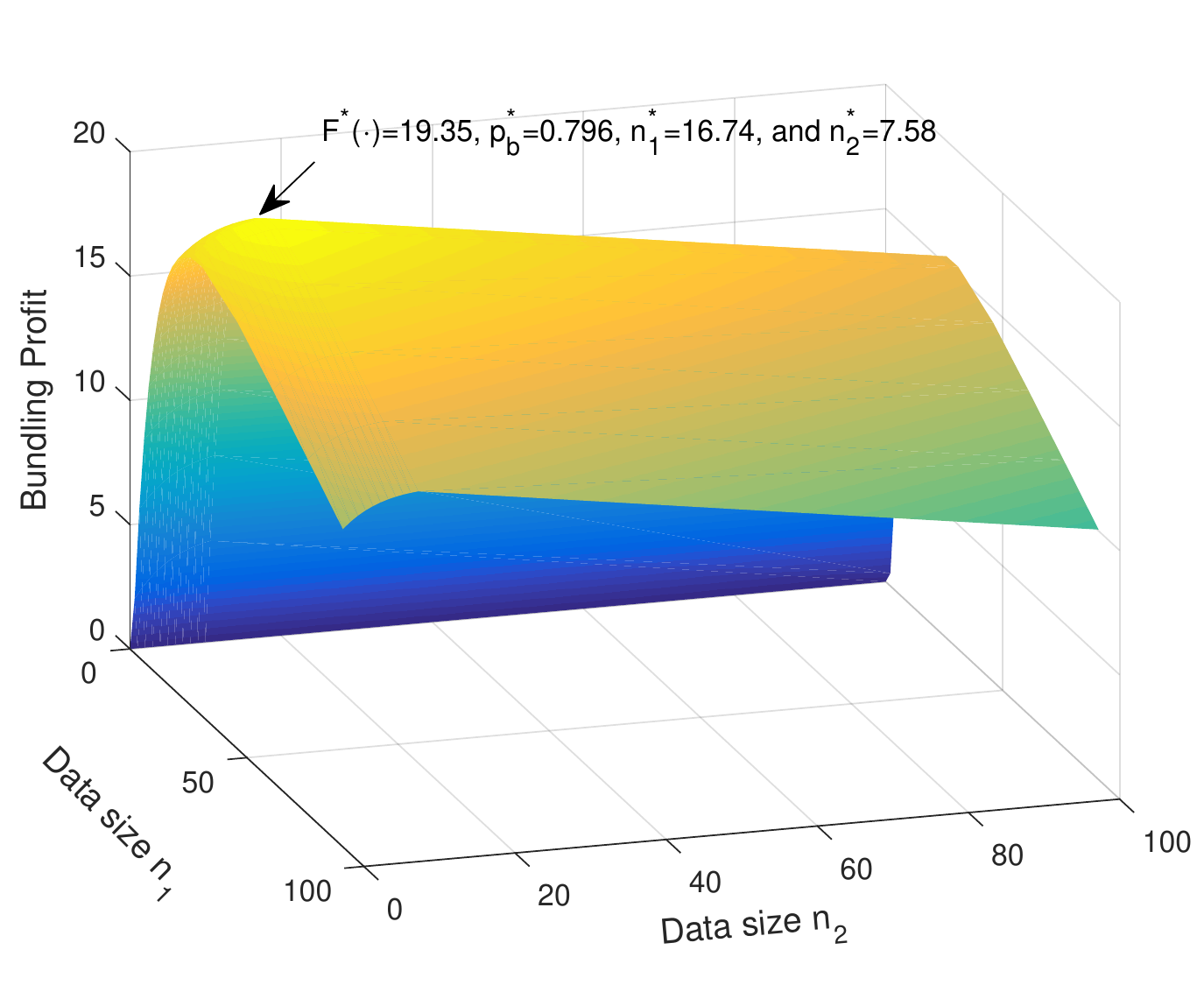}
				\par\end{centering}
		}
		\subfloat[]{\begin{centering}
				\includegraphics[width=0.95\columnwidth,trim=0cm 0.5cm 0cm 0.5cm]{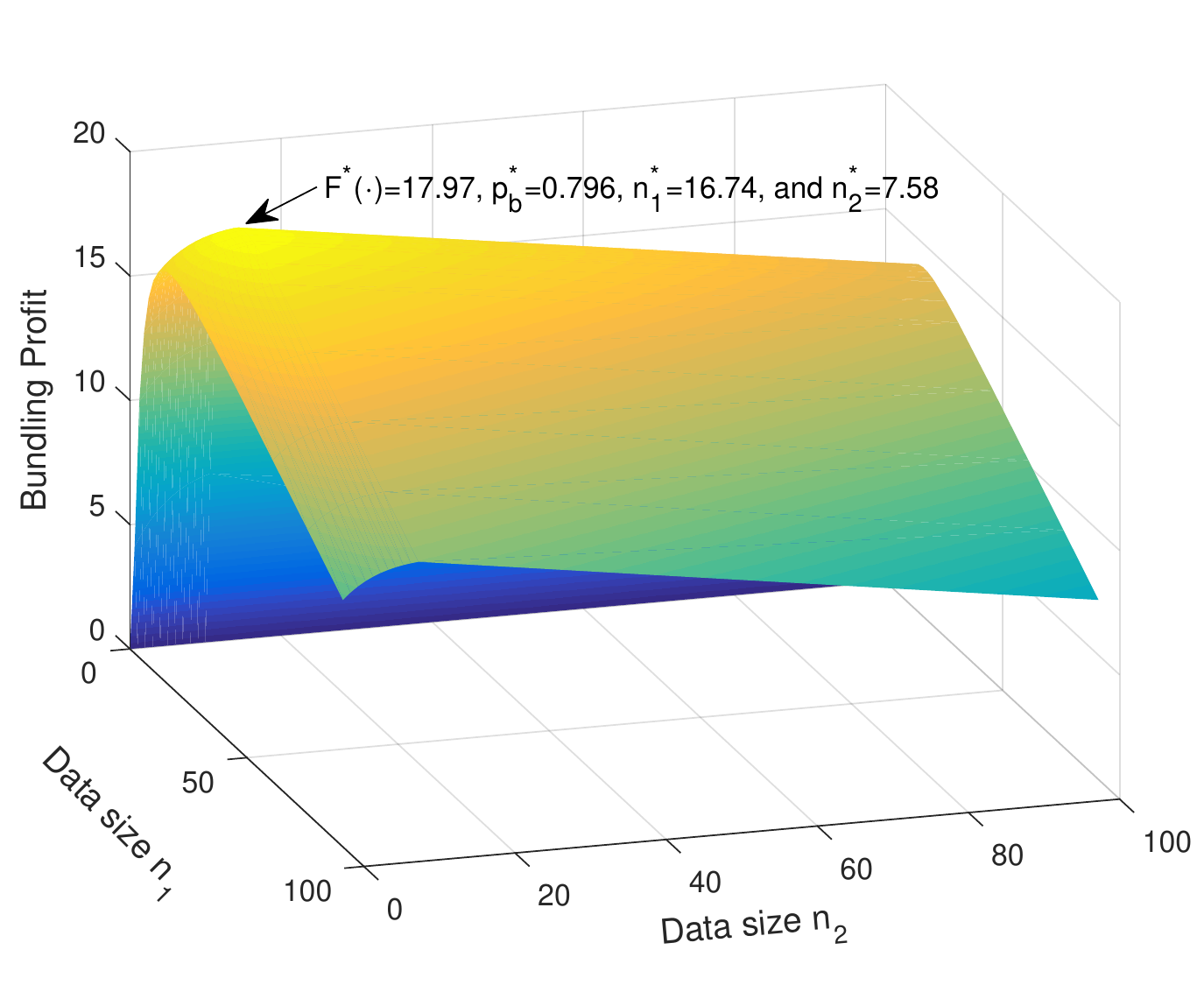}
				\par\end{centering}
		}
		\par\end{centering}
	
	\caption{Profit from the bundled service under varied data sizes ($n_{1}$ and $n_{2}$). (a)-(d) are for Cases~1-4, respectively.\label{fig:bundle_profit_size}}
\end{figure*}

\subsection{Service Bundling}

In the following, we consider the marketing strategy of grouping Services~1 and 2 into one bundled service forming a coalition $\mathcal{K}$. Deep learning is used for machine learning in both services resulting in quality functions as shown in Figure~\ref{fig:data_utility}. Specifically, the service quality parameters are $\alpha_{11}=0.884$, $\alpha_{21}=0.59$ , and $\alpha_{31}=0.114$ for Services~1, while the quality parameters are $\alpha_{12}=0.82$, $\alpha_{22}=0.069$ , and $\alpha_{32}=0.142$ for Service~2. We use the analysis of Section~\ref{sec:service_bundling} for service bundling and pricing.

\subsubsection{Profit Maximization\label{sub:profit_maximization}}

The profits $F_{\mathcal{K}}\left(p_{b},n_{1},n_{2}\right)$ of the bundled service is plotted in Figure~\ref{fig:bundle_profit_size} which correspond to the four demand cases of (\ref{eq:demand_case_1})-(\ref{eq:demand_case_4}). The maximum profit is achieved by using Case~1 ($C1:p_{b}\leq q_{1}$ and $C2:p_{b}\leq q_{2}$) with the closed-form solutions defined in (\ref{eq:bundling_sol_n1})-(\ref{eq:bundling_sol_lagrange}). The optimal solutions are $n_{1}^{*}=19.29$, $n_{2}^{*}=7.01$, $p_{b}^{*}=0.658$ which generate an optimal profit of $F_{\mathcal{K}}\left(p_{b}^{*},n_{1}^{*},n_{2}^{*}\right)=19.67$.

\begin{figure}
	\begin{centering}
		\subfloat[\label{fig:market_decision_1}]{\begin{centering}
				\includegraphics[width=0.9\columnwidth,trim=0cm 0.5cm 0cm 0cm]{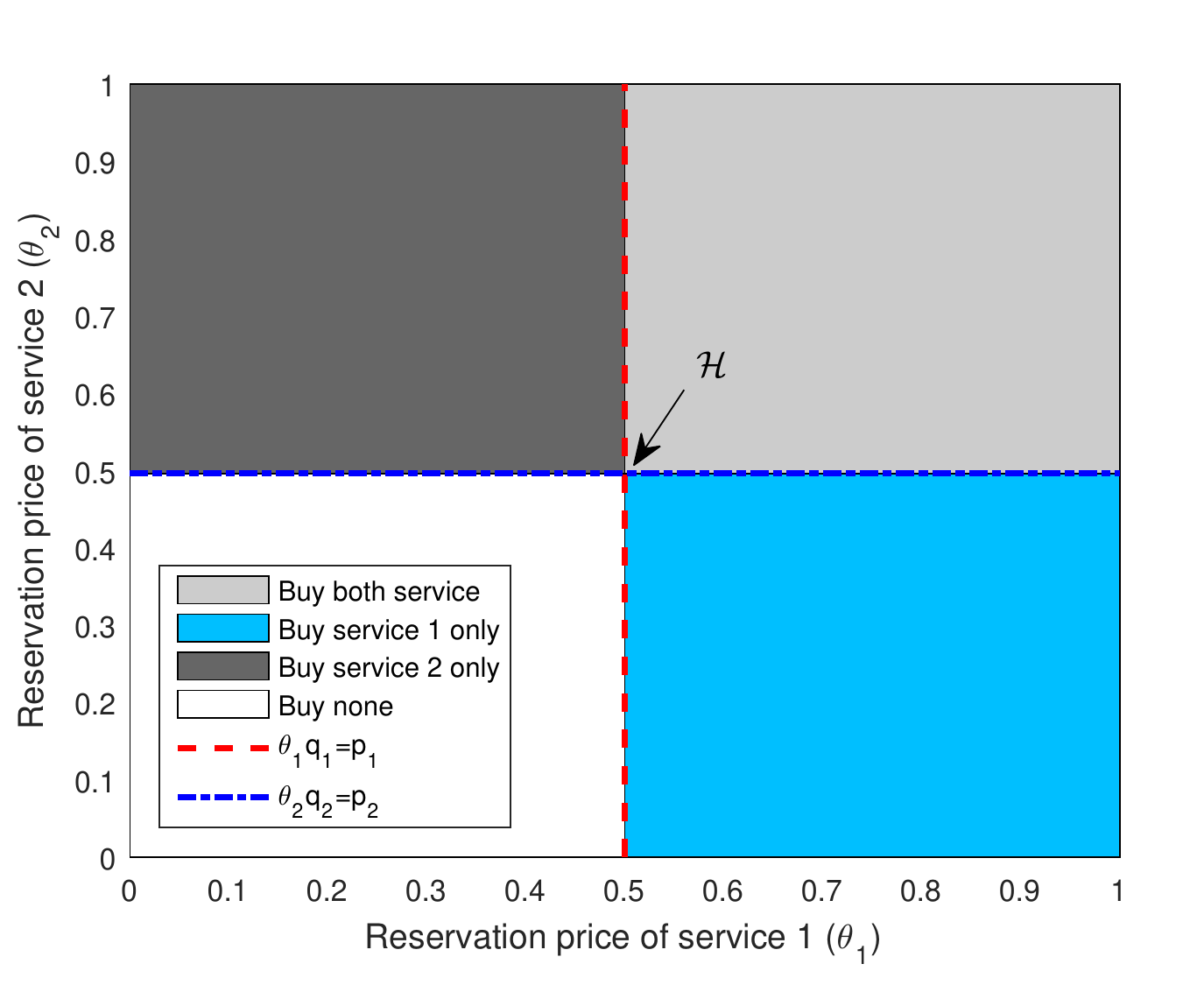}
				\par\end{centering}
		}
		
		\subfloat[\label{fig:market_decision_2}]{\begin{centering}
				\includegraphics[width=0.9\columnwidth,trim=0cm 0.5cm 0cm 1cm]{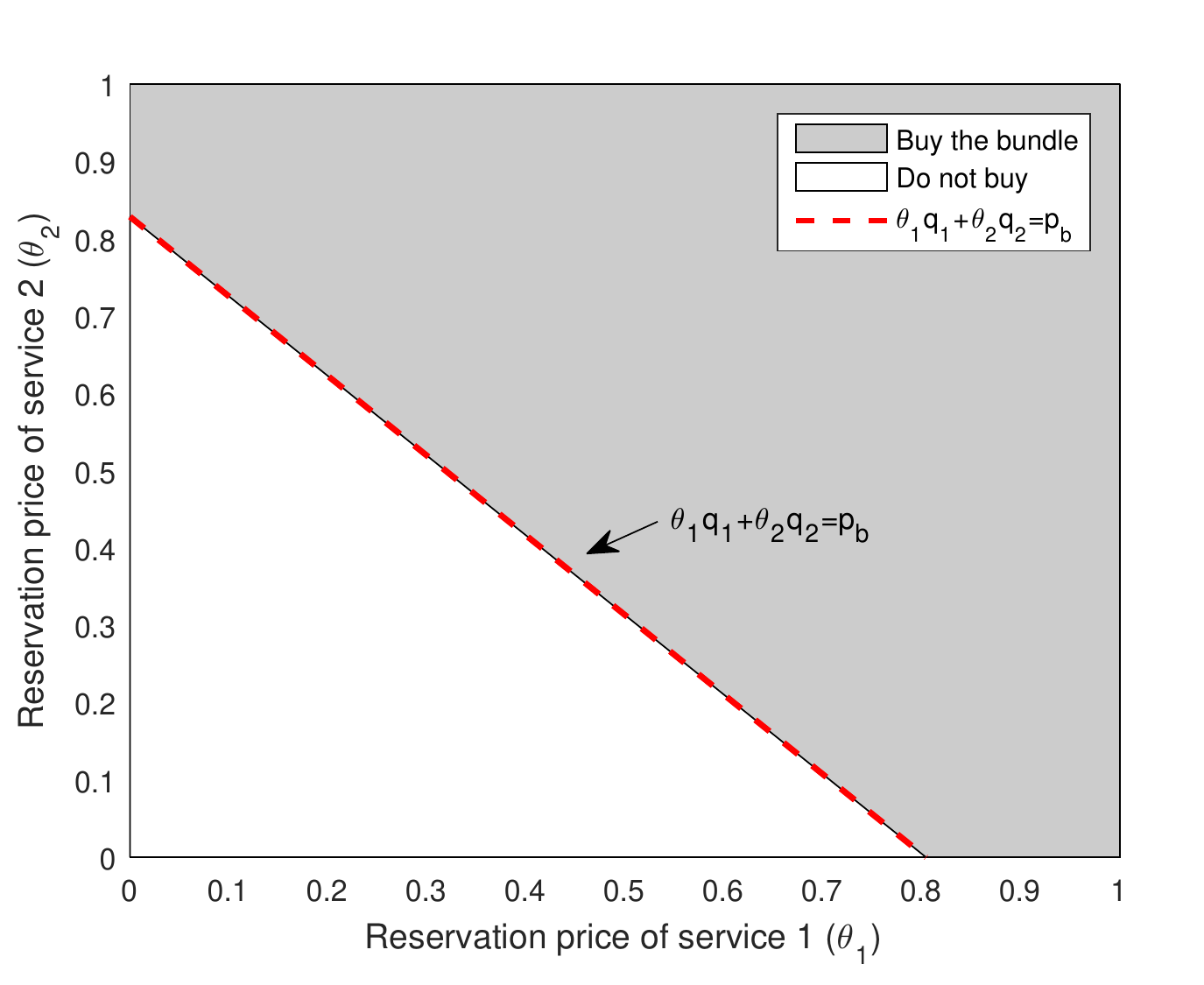}
				\par\end{centering}
			
		}
		\par\end{centering}
	
	\caption{Market equilibrium under different customer reservations. (a)~Separate selling of the two services with subscription fees of $p_{s1}^{*}=0.41$ and $p_{s2}^{*}=0.39$. (b)~Selling the two services as a bundle with bundle subscription fee of $p_{b}=0.659$.\label{fig:market_decision}}
\end{figure}

Figure~\ref{fig:market_decision} illustrates the market equilibrium of selling Services~1 and 2 separately or as one bundle under coalition $\mathcal{K}$. Using (\ref{eq:seperate_sol_n}), (\ref{eq:eq:seperate_sol_ps}), and (\ref{eq:seperate_profit}) to analyze the separate sales, we find that the optimal subscription fee and profit values of Services~1 and 2 are $p_{s1}^{*}=0.41$, $F_{1}\left(n_{1}^{*},p_{s1}^{*}\right)=8.31$, $p_{s2}^{*}=0.39$, and $F_{2}\left(n_{2}^{*},p_{s2}^{*}\right)=9.58$, respectively.
\begin{itemize}
	\item For separate sales (Figure~\ref{fig:market_decision_1}), the optimal market equilibrium is shown as the vertical $\theta_{1}q_{1}=p_{s1}^{*}=0.41$ and horizontal \textbf{$\theta_{2}q_{2}=p_{s2}^{*}=0.39$} lines for Services~1 and 2, respectively. There are four decision areas centered by the point $\mathcal{H}\equiv\left(\frac{p_{s1}^{*}}{q_{1}},\frac{p_{s2}^{*}}{q_{2}}\right)$. Firstly, the customers with reservation prices that lie to the northeast of $\mathcal{H}$ will subscribe to both services. Secondly, the customers with reservation prices that lie to the southeast of $\mathcal{H}$ will subscribe to Service~1 only, Thirdly, the customers with reservation prices that lie to the northwest of $\mathcal{H}$ will subscribe to Service~2 only, Finally, the customers with reservation prices that lie to the southwest of $\mathcal{H}$ will not subscribe to any service as their evaluations of both service are below the market equilibrium values.
	\item Figure~\ref{fig:market_decision_2} shows that customers with reservation prices that lie above the line $\theta_{1}q_{1}+\theta_{2}q_{2}=p_{b}^{*}$ will decide to subscribe to the bundled service. On the other hand, the customers with reservation prices that lie below the line $\theta_{1}q_{1}+\theta_{2}q_{2}=p_{b}^{*}$ will not subscribe to the bundle.
	
\end{itemize}
We can also observe that the customers will be interested in buying the two services as one bundled service at the discounted subscription price $p_{b}^{*}<p_{s1}^{*}+p_{s2}^{*}$. From the provider's perspectives, the bundled service helps in maximizing their profits $F_{\mathcal{K}}\left(p_{b}^{*},n_{1}^{*},n_{2}^{*}\right)>F_{1}\left(n_{1}^{*},p_{s1}^{*}\right)+F_{2}\left(n_{2}^{*},p_{s2}^{*}\right)$. This profit is shared among the two service providers as will be discussed later in Section~\ref{sub:profit_sharing}.

\subsubsection{The Impact of Data Cost}

\begin{figure}
	\begin{centering}
		\includegraphics[width=1.0\columnwidth,trim=0cm 0.5cm 0cm 0cm]{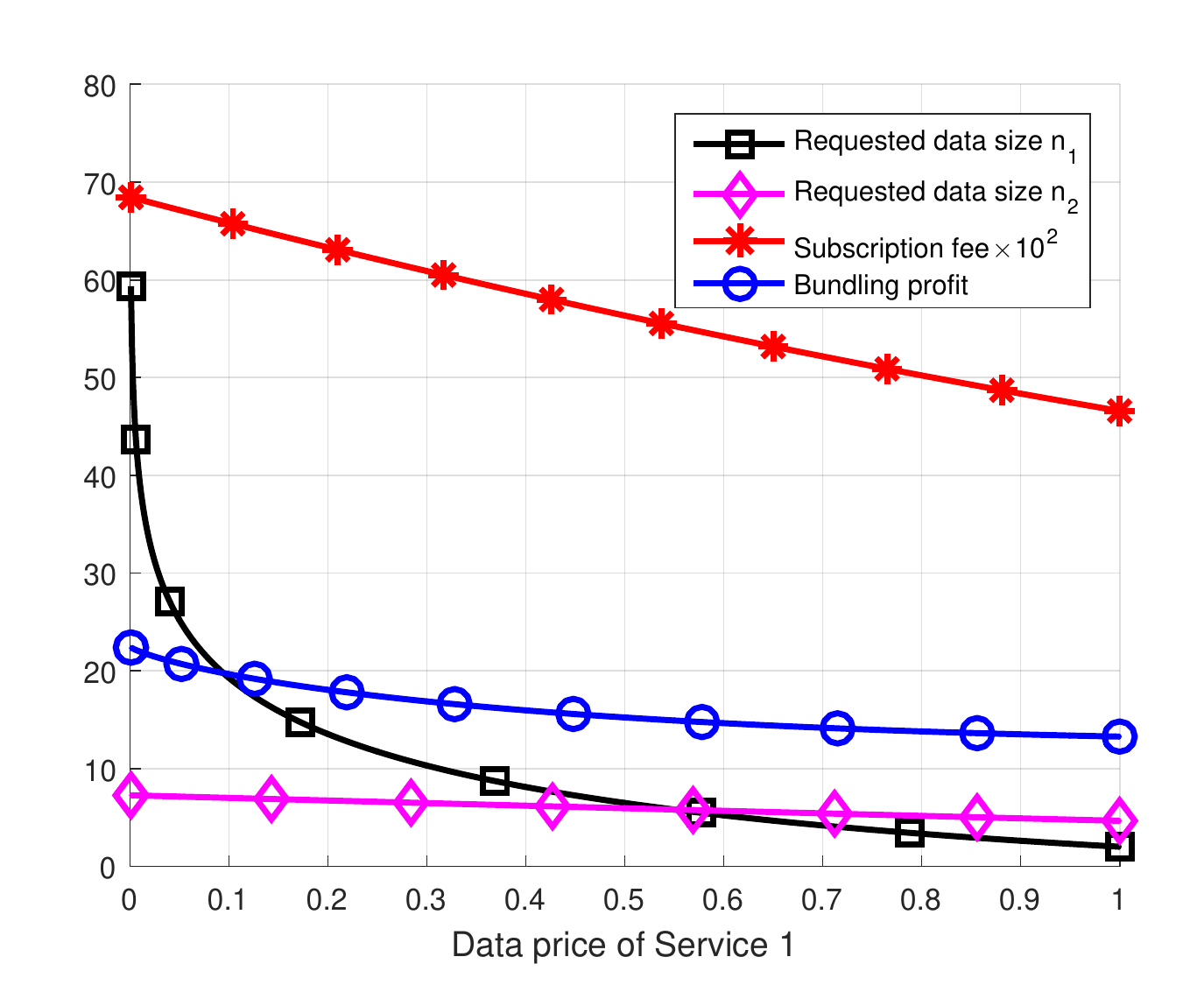}
		\par\end{centering}
	
	\caption{Optimal requested data sizes $n_{1}^{*}$ and $n_{2}^{*}$, bundle subscription fee $p_{b}^{*}$, and bundling profit $F_{\mathcal{K}}^{*}\left(\cdot\right)$ of service providers under varied data prices of Service~1 $c_{1}$.\label{fig:bundle_data_price}}
\end{figure}

While fixing the data price of Service~2 ($c_{2}=0.05$), we have varied the data price of Service~1 and observed the market parameters of the bundled service as in Figure~\ref{fig:bundle_data_price}. There are several important observations that can be highlighted. The optimal requested data sizes $n_{1}^{*}$ and $n_{2}^{*}$, bundle subscription fee $p_{b}^{*}$, and bundling profit $F_{\mathcal{K}}\left(p_{b}^{*},n_{1}^{*},n_{2}^{*}\right)$ decrease as the data price $c_{1}$ increases. Specifically, when the data price $c_{1}$ is high, the cooperative providers will try to minimize the total cost by buying less data for Service~1 $n_{1}^{*}$. Likewise, this will decrease the ability of buying data for Service~2 $n_{2}^{*}$ which will be slightly decreased. Finally, to maintain the market equilibrium, the subscription fee $p_{b}^{*}$ of the service bundle should be also decreased for high values of data price $c_{1}$. This inverse correlation can be deduced from (\ref{eq:bundle_decision}) where the customers consider the qualities of Services~1 and 2 as well as the bundle subscription fee when making their subscription decisions.

\subsubsection{The Impact of Customer Base Size}

\begin{figure}
	\begin{centering}
		\includegraphics[width=1.0\columnwidth]{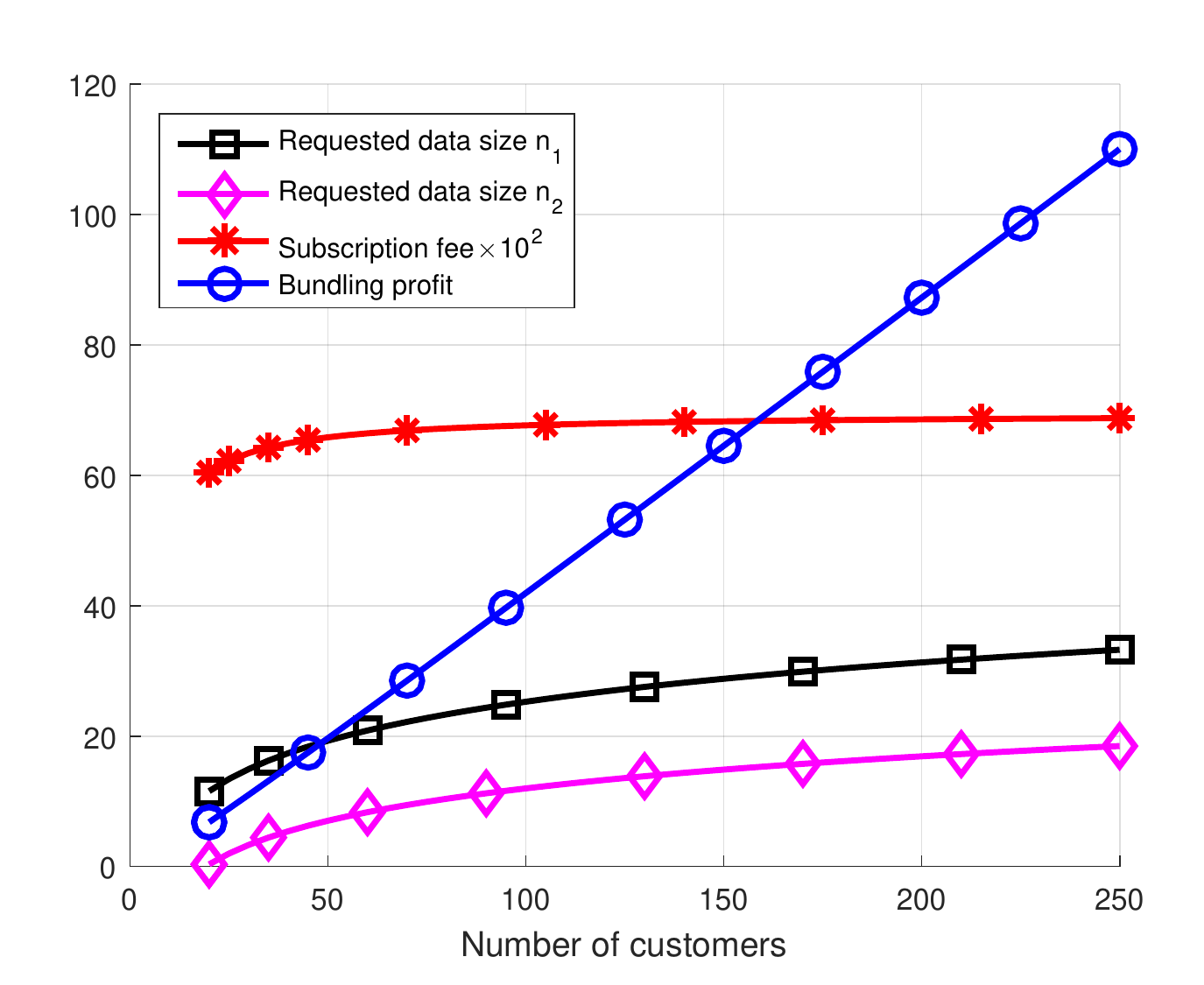}
		\par\end{centering}
	
	\caption{Optimal requested data sizes $n_{1}^{*}$ and $n_{2}^{*}$, bundle subscription fee $p_{b}^{*}$, and bundling profit $F_{\mathcal{K}}^{*}\left(\cdot\right)$ of service providers under varied sizes of the customer base $M$.\label{fig:bundle_customers}}
\end{figure}

Figure~\ref{fig:bundle_customers} shows that the optimal requested data sizes $n_{1}^{*}$ and $n_{2}^{*}$, subscription fee $p_{b}^{*}$, and bundling profit $F_{\mathcal{K}}\left(p_{b}^{*},n_{1}^{*},n_{2}^{*}\right)$ are proportional to the number of customers $M$. In particular, an increased number of customers $M$ results in higher values of $n_{1}^{*}$, $n_{2}^{*}$, $p_{b}^{*}$, and $F_{\mathcal{K}}\left(p_{b}^{*},n_{1}^{*},n_{2}^{*}\right)$. As any other demand relationship in economics, the subscription fee and profit increase for high demands by customers. Similarly, the service providers will buy more data to provide better quality services as more customers would be interested in buying the service. 

\subsection{Profit Sharing of the Service Bundle\label{sub:profit_sharing}}

We next analyze the profit sharing among the two service providers forming the bundled service.

\subsubsection{Payoff Allocation}

\begin{figure}
	\begin{centering}
		\includegraphics[width=0.9\columnwidth]{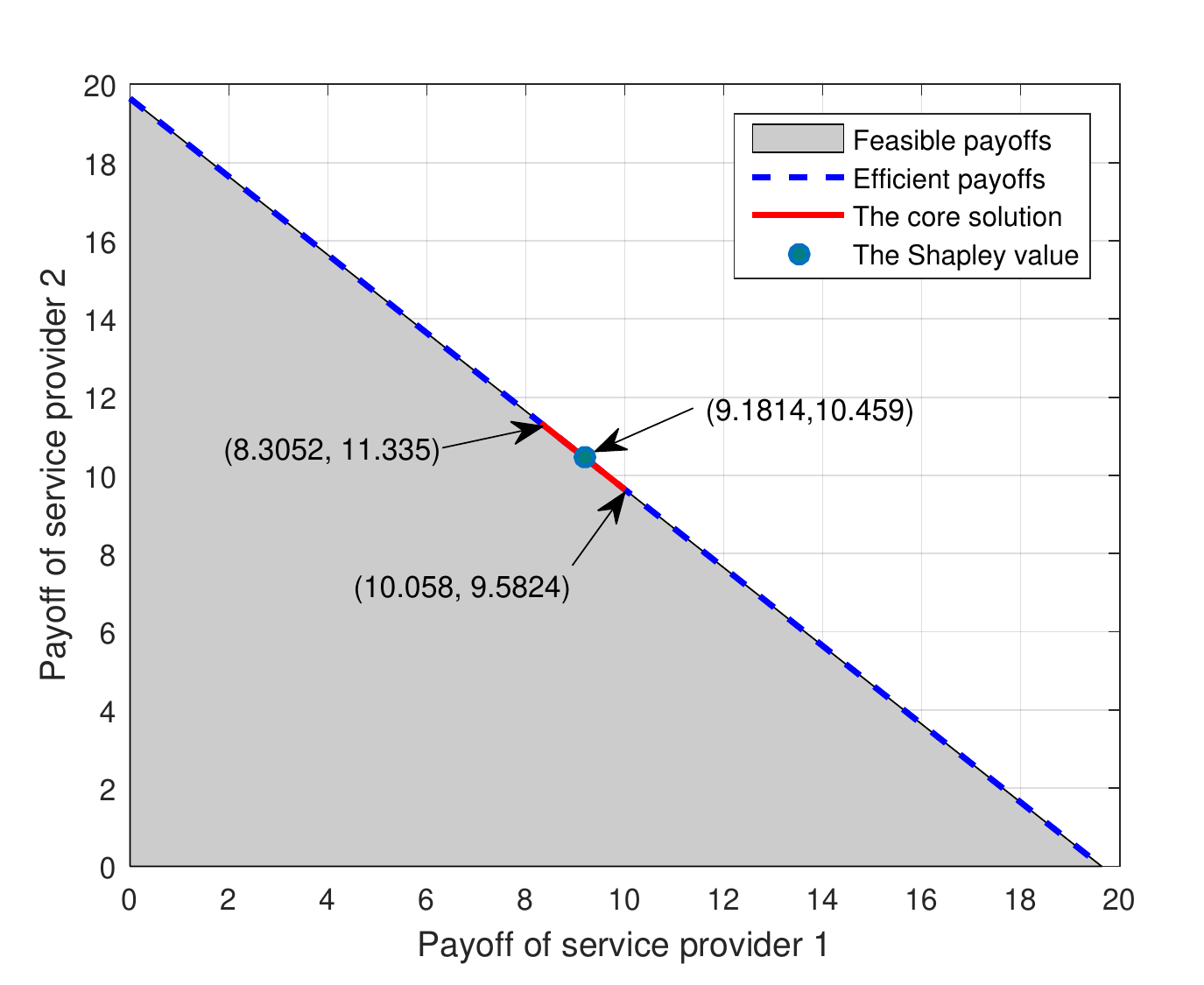}
		\par\end{centering}
	\caption{Profit sharing of the bundled service.\label{fig:payoffs_sol}}	
\end{figure}

Recall from Section~\ref{sub:profit_maximization} that the bundling profit is $F_{\mathcal{K}}\left(p_{b}^{*},n_{1}^{*},n_{2}^{*}\right)=19.67$, while the profits of separate selling are $F_{1}\left(n_{1}^{*},p_{s1}^{*}\right)=8.31$ and $F_{2}\left(n_{2}^{*},p_{s2}^{*}\right)=9.58$ of Services~1 and 2, respectively. Figure~\ref{fig:payoffs_sol} shows four important solution concepts. The first solution (gray shaded area) gives the feasible payoffs. The second solution gives ``efficient payoffs'' which satisfy the constraint $\varphi_{1}+\varphi_{2}=F_{\mathcal{K}}\left(p_{b}^{*},n_{1}^{*},n_{2}^{*}\right)$. However, this solution does not guarantee that the individual payoff of each provider resulting from the bundling strategy is higher than that of the separate service selling. Therefore, a provider may reject his payoff and decide to sell his service separately. The third solution is the core which is found as in (\ref{eq:core_sol}). This solution guarantees that the individual payoffs of providers are higher with bundling compared to the separate selling of services. The final solution is the Shapley value as in (\ref{eq:shapley_sol}) which gives fair payoffs for the cooperative providers in $\mathcal{K}$ based on their individual contribution to the service bundle.

\subsubsection{The Impact of Individual Contributions on Payoffs}

\begin{figure}
	\begin{centering}
		\includegraphics[width=0.9\columnwidth, trim=0.5cm 0.5cm 0.5cm 0cm]{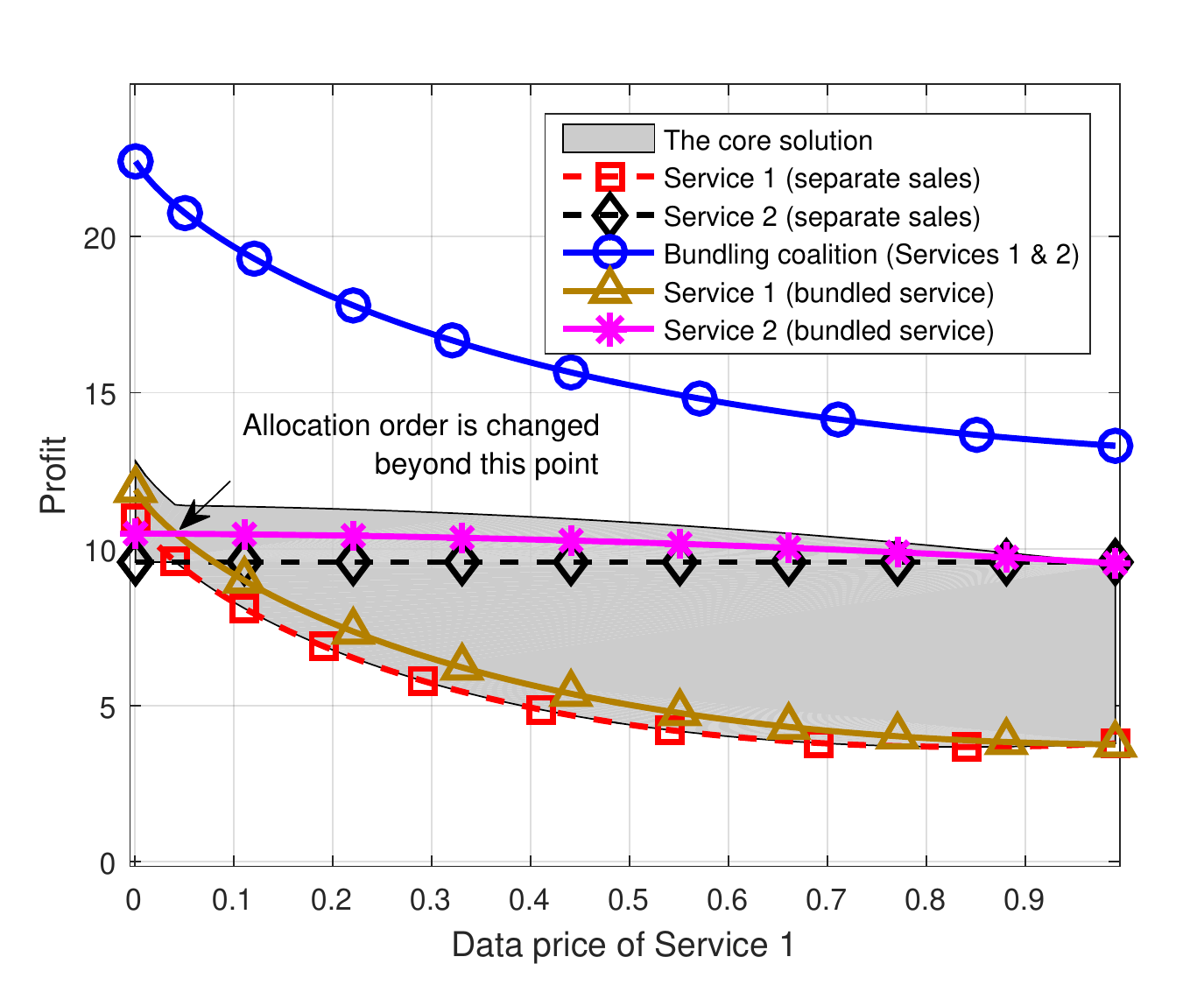}
		\par\end{centering}
	
	\caption{Profit of Services~1 and 2 under varied data prices of Service~1 $c_{1}$.\label{fig:payoffs_data_price}}
\end{figure}

In Figure~\ref{fig:payoffs_data_price}, we further analyze the fair allocation of payoffs with the Shapley concept. In particular, we consider the case when the data price of Service~1 is varied while fixing the data price of Service~2 ($c_{2}=0.05$). The payoff allocations by the Shapley value are always inside the core solution area (the gray shaded area). Firstly, we compare the profit of selling Service~1 and 2 separately or as a bundled service. The payoff allocations to each service by the Shapley value are higher than those of the separate sales. In the separate selling, changing the price of one service does not change the profit of the other service. Secondly, the payoff allocations in the bundled service reflect the contributions of an individual service to the profit of the bundled service. When the data price of a service is high, the payoff of that particular service will be fairly decreased due to the high cost introduced in building the bundle. This fair allocation of payoff in the Shapley concept provides an incentive for cooperative service providers to search for the best deals and vendors of data. Thirdly, it can be noted that the high data cost of Service~1 reduces the incentive of Service~2 in the bundle formation as the residual margin between the profit of separate and bundled sales diminishes and becomes zero at $c_{1}=1$.

\begin{figure}
	\begin{centering}
		\includegraphics[width=0.9\columnwidth, trim=0.5cm 0.5cm 0.5cm 0cm]{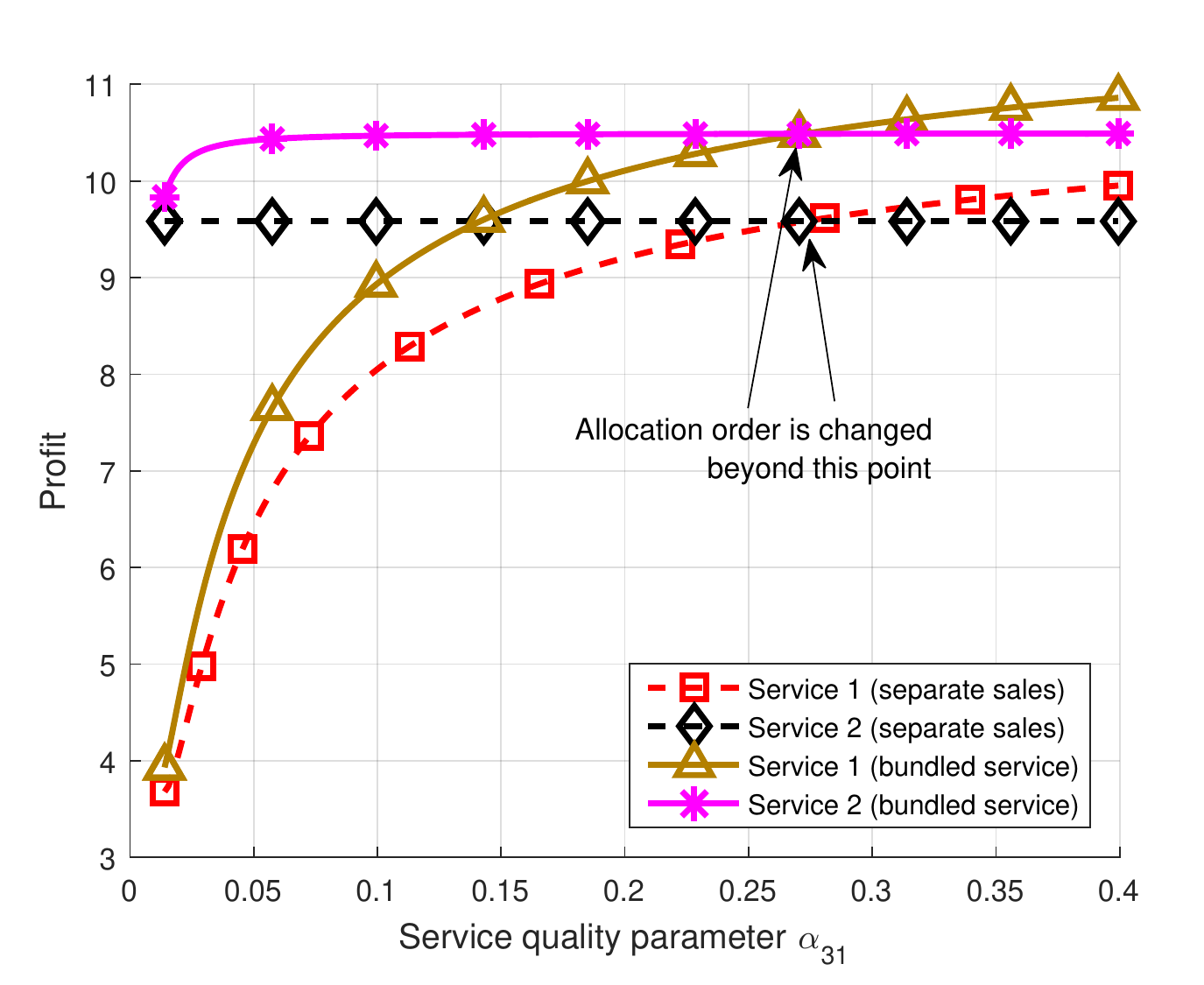}
		\par\end{centering}
	
	\caption{Profit of Services~1 and 2 under varied quality of Service~1 $\alpha_{31}$.\label{fig:payoffs_utility_param}}
\end{figure}

Finally, Figure~\ref{fig:payoffs_utility_param} shows that higher quality of one service increases the profit of the other cooperative service within the bundle, and vice versa. When the quality of Service~1 is increased (high values of $\alpha_{31}$) while fixing the quality of Service~2, the payoff allocations to Services~1 and 2 increase. However, it is observed that the profit increase of Service~1 is greater than that of Service~2. In the separate selling case, Service~2 will not benefit from enhancing the quality of Service~1. Therefore, there is a profit incentive for service providers to look for high quality services to cooperate with when forming bundled services.

\section{Conclusion\label{sec:conclusion}}

In this paper, we have developed IoT market models and optimal pricing schemes for the standalone and bundled sales of machine learning-based IoT services. The proposed optimizations maximize the profit of service providers by deciding the optimal data sizes that should be bought from data vendors and the optimal subscription fees at which the services should be offered to customers. Bundling IoT services has been shown as an effective marketing strategy for maximizing the profits of providers. We have observed that service providers have a profit incentive in searching for the best quality services to form a bundling coalition. Finally, we have presented a profit sharing models for allocating the profit payoffs among cooperative service providers. 

In this paper, we have assumed that the service providers are economically rational, i.e.,~they take decisions that maximize their revenues. Future work can improve our optimizations by exploring the possible competition among service providers. Moreover, our profit maximization models can be extended to support  complementary and substitute services.

\section*{Acknowledgment}

This work was supported in part by the Australian Research Council (ARC) under grant DE200100863. It was also partially supported by Singapore NRF National Satellite of Excellence, Design Science and Technology for Secure Critical Infrastructure NSoE DeST-SCI2019-0007, A*STAR-NTU-SUTD Joint Research Grant Call on Artificial Intelligence for the Future of Manufacturing RGANS1906, WASP/NTU M4082187 (4080), Singapore MOE Tier 1 2017-T1-002-007 RG122/17, MOE Tier 2 MOE2014-T2-2-015 ARC4/15, Singapore NRF2015-NRF-ISF001-2277, Singapore EMA Energy Resilience NRF2017EWT-EP003-041, US MURI AFOSR MURI 18RT0073, NSF EARS-1839818, CNS1717454, CNS-1731424, CNS-1702850, and CNS-1646607.

\bibliographystyle{ieeetr}
\bibliography{paper}

\section*{Biographies}
\begin{IEEEbiography}[{\includegraphics[width=1in,height=1.25in,clip,keepaspectratio]{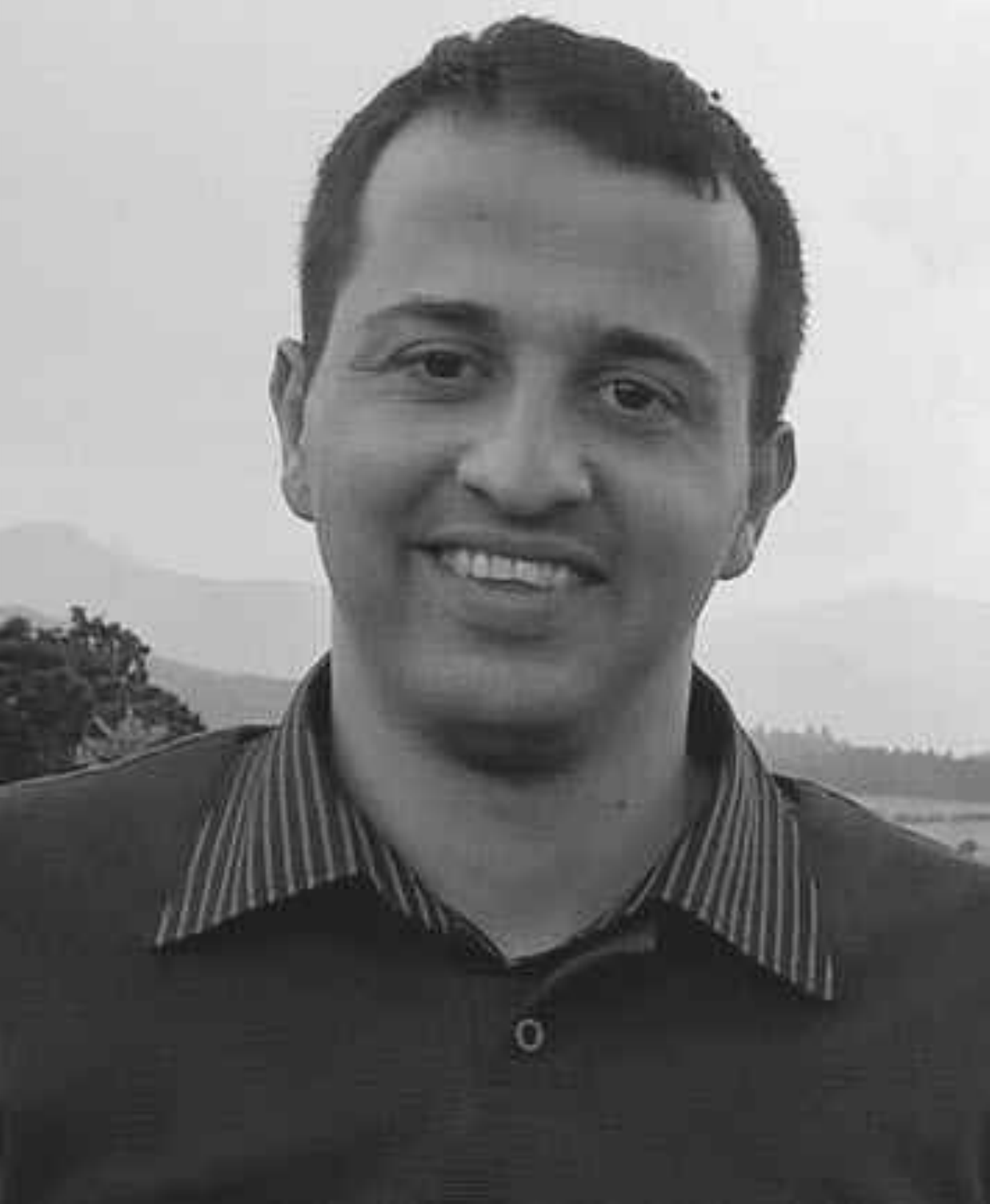}}]
{Mohammad Abu Alsheikh}
(mabualsh@ieee.org) [S'14--M'17] is an Assistant Professor and ARC DECRA Fellow at the University of Canberra, Australia. He was a Postdoctoral Researcher at Massachusetts Institute of Technology, USA. He designs and creates novel IoT systems that leverage both machine learning and convex optimization with applications in people-centric sensing, human activity recognition, and smart cities. His doctoral research at Nanyang Technological University, Singapore was focused on optimizing the data collection in wireless sensor networks. After graduating with a B.Eng. degree in computer systems from Birzeit University, Palestine, he worked as a software engineer at a digital advertising start-up and Cisco.
\end{IEEEbiography}

\begin{IEEEbiography}[{\includegraphics[width=1in,height=1.25in,clip,keepaspectratio]{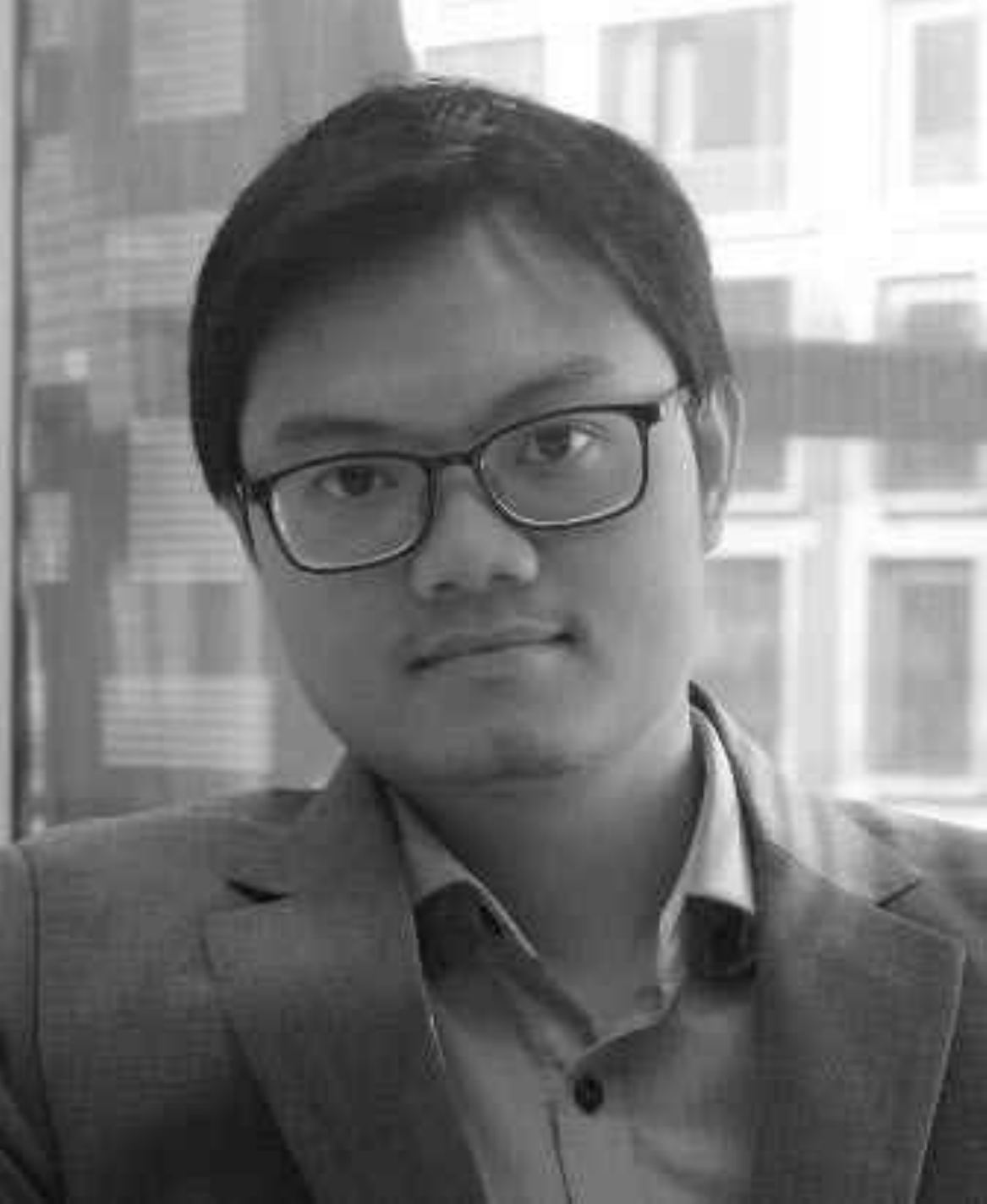}}]
{Dinh Thai Hoang}
(hoang.dinh@uts.edu.au) [M'16] is a faculty member with the School of Electrical and Data Engineering, the University of Technology Sydney, Australia. He received his Ph.D. in Computer Science and Engineering from the Nanyang Technological University, Singapore, in 2016. His research interests include emerging topics in wireless communications and networking such as ambient backscatter communications, cognitive radios, wireless energy harvesting, IoT, mobile edge and 5G networks. He has been serving as an active member of technical program committees and a reviewer for many IEEE Transactions, Journals, Magazines, and Conferences. He received the best reviewer award from IEEE Transactions of Wireless Communications and he is currently an editor of IEEE Wireless Communications Letters and IEEE Transactions on Cognitive Communications and Networking.
\end{IEEEbiography}

\begin{IEEEbiography}[{\includegraphics[width=1in,height=1.25in,clip,keepaspectratio]{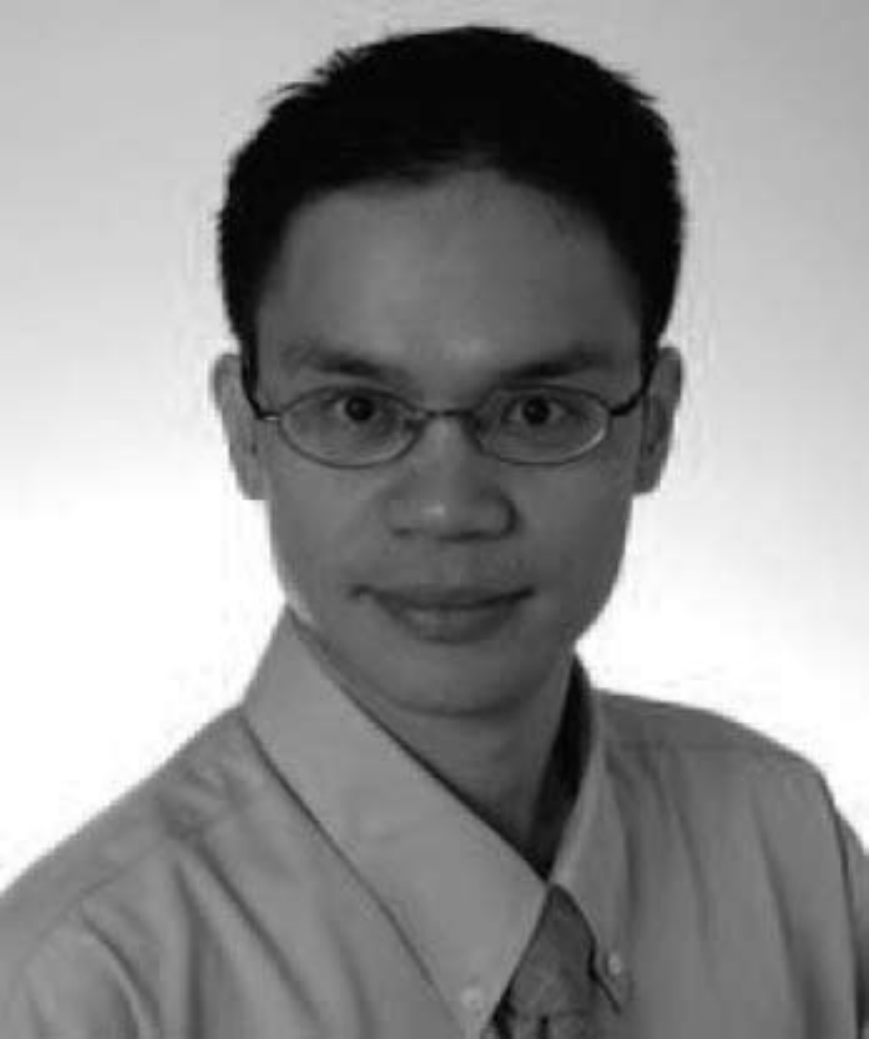}}]
{Dusit Niyato}
(dniyato@ntu.edu.sg) [M'09--SM'15--F'17] is currently a Professor in the School of Computer Science and Engineering, at Nanyang Technological University, Singapore. He received B.Eng. from King Mongkuts Institute of Technology Ladkrabang (KMITL), Thailand in 1999 and Ph.D. in Electrical and Computer Engineering from the University of Manitoba, Canada in 2008. His research interests are in the area of energy harvesting for wireless communication, Internet of Things (IoT) and sensor networks.
\end{IEEEbiography}
\vfill

\begin{IEEEbiography}[{\includegraphics[width=1in,height=1.25in,clip,keepaspectratio]{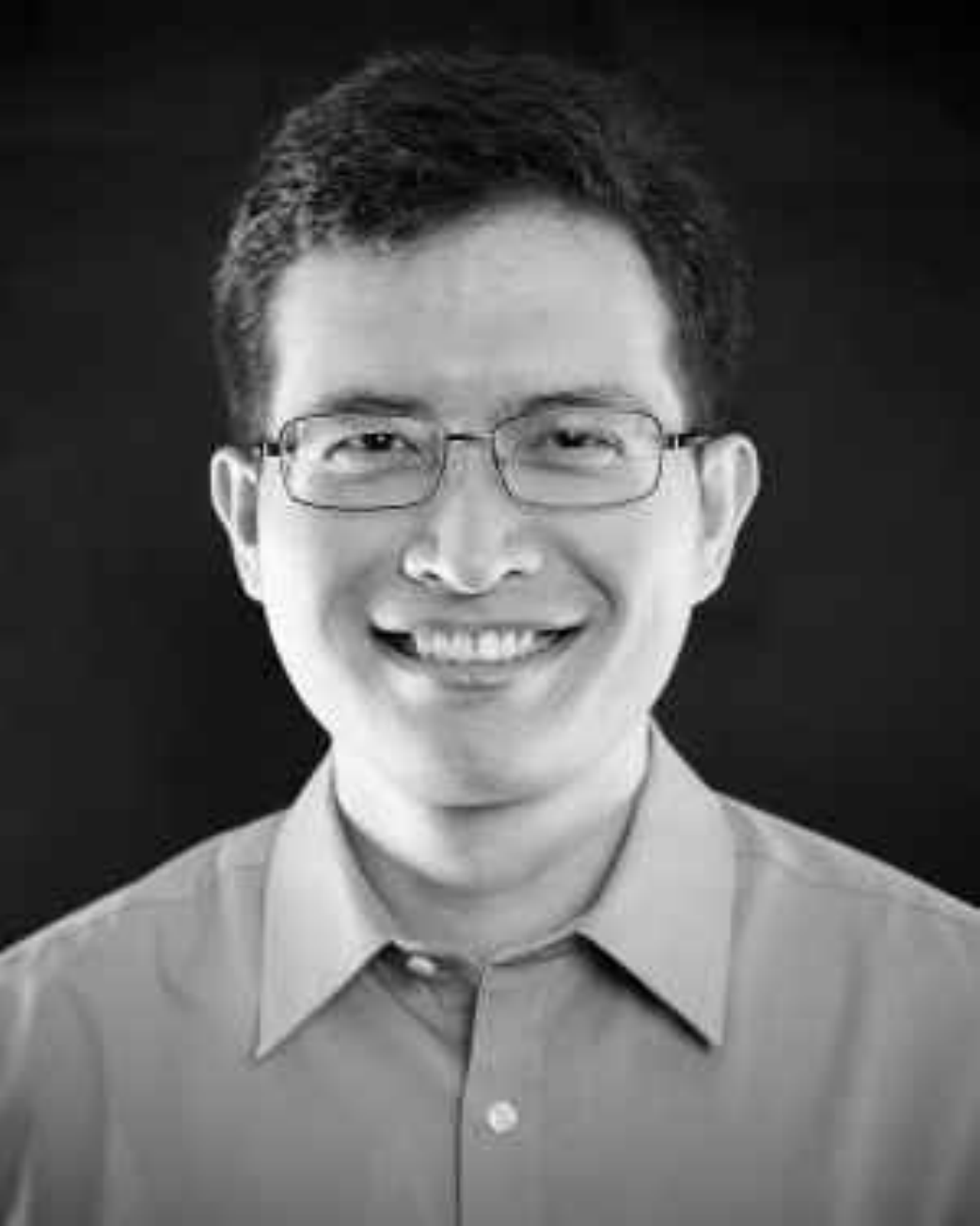}}]
{Derek Leong} 
(derekleong@alumni.cmu.edu) received the B.S. degree in electrical and computer engineering from Carnegie Mellon University in 2005, and the M.S. and Ph.D. degrees in electrical engineering from the California Institute of Technology in 2008 and 2013, respectively. He was a Scientist with the Smart Energy and Environment cluster at the Institute for Infocomm Research~(I\textsuperscript{2}R), A*STAR, Singapore. His research and development interests include distributed systems, sensor networks, smart cities, and the Internet of Things.
\end{IEEEbiography}

\begin{IEEEbiography}[{\includegraphics[width=1in,height=1.25in,clip,keepaspectratio]{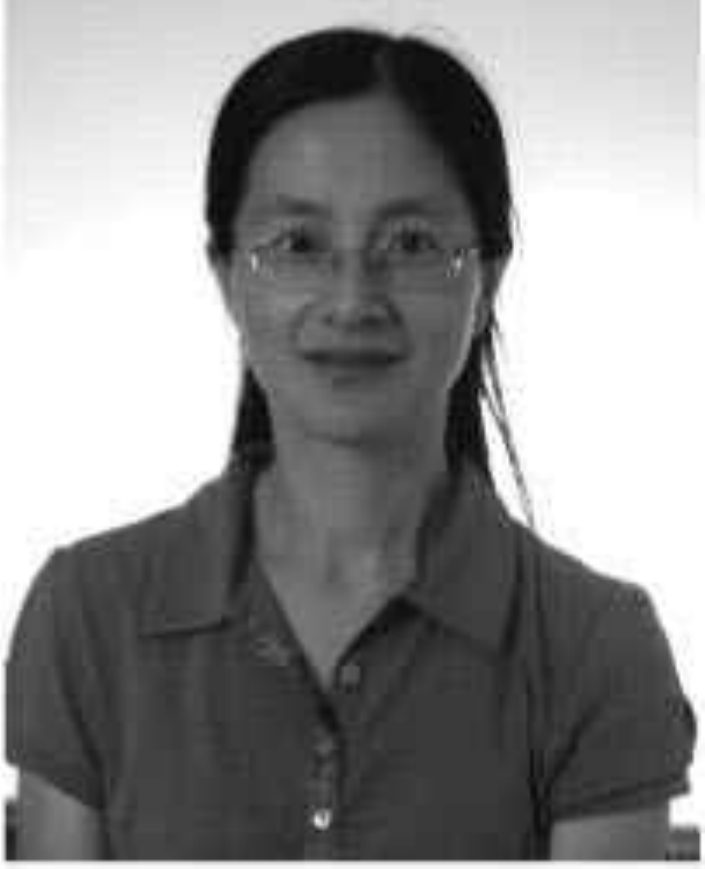}}]
{Ping Wang}
(wangping@ntu.edu.sg) [M'08--SM'15] received the Ph.D. degree in electrical engineering from University of Waterloo, Canada, in 2008. Currently she is an Associate Professor in the Lassonde School of Engineering, York University, Canada. Her current research interests include resource allocation in multimedia wireless networks, cloud computing, and smart grid. She was a corecipient of the Best Paper Award from IEEE Wireless Communications and Networking Conference~(WCNC) 2012 and IEEE International Conference on Communications~(ICC) 2007.
\end{IEEEbiography}

\begin{IEEEbiography}[{\includegraphics[width=1in,height=1.25in,clip,keepaspectratio]{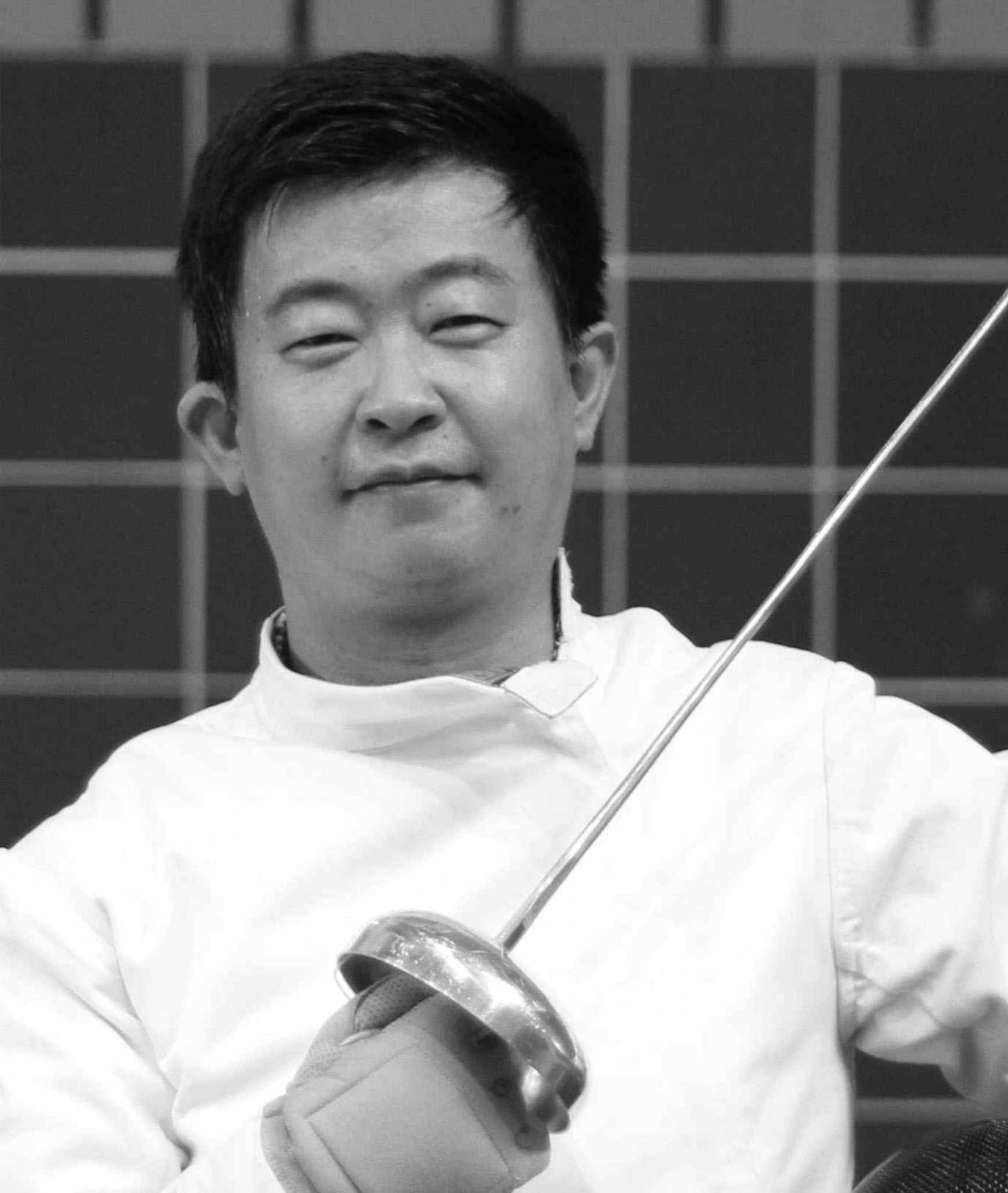}}]
{Zhu Han}
(zhan2@uh.edu) [S'01--M'04--SM'09--F'14] received the B.S. degree in electronic engineering from Tsinghua University, in 1997, and the M.S. and Ph.D. degrees in electrical engineering from the University of Maryland, College Park, in 1999 and 2003, respectively. From 2000 to 2002, he was an R\&D Engineer of JDSU, Germantown, Maryland. From 2003 to 2006, he was a Research Associate at the University of Maryland. From 2006 to 2008, he was an assistant professor in Boise State University, Idaho. Currently, he is a Professor in Electrical and Computer Engineering Department as well as Computer Science Department at the University of Houston, Texas. His research interests include wireless resource allocation and management, wireless communications and networking, game theory, wireless multimedia, security, and smart grid communication. Dr. Han received an NSF Career Award in 2010, the Fred W. Ellersick Prize of the IEEE Communication Society in 2011, the EURASIP Best Paper Award for the Journal on Advances in Signal Processing in 2015, several best paper awards in IEEE conferences, and is currently an IEEE Communications Society Distinguished Lecturer.
\end{IEEEbiography}
\vfill

\end{document}